\documentclass[aps,pre,twocolumn,superscriptaddress,nofootinbib]{revtex4}

\usepackage{graphicx,amssymb,amsfonts,amsmath,chemarr,color,textcomp}

\newcommand{\beq}{\begin{equation}}
\newcommand{\eeq}{\end{equation}}
\newcommand{\beqn}{\begin{eqnarray}}
\newcommand{\eeqn}{\end{eqnarray}}
\newcommand{\elabel}[1]{\label{eq:#1}}
\newcommand{\eref}[1]{Eqn.\ \ref{eq:#1}}
\newcommand{\erefs}[2]{Eqns.\ \ref{eq:#1} and \ref{eq:#2}}

\newcommand{\erefn}[2]{Eqns.\ \ref{eq:#1}-\ref{eq:#2}}
\newcommand{\flabel}[1]{\label{fig:#1}}
\newcommand{\fref}[1]{Fig.\ \ref{fig:#1}}

\newcommand{\slabel}[1]{\label{sec:#1}}
\newcommand{\sref}[1]{Sec.\ \ref{sec:#1}}
\newcommand{\alabel}[1]{\label{app:#1}}
\newcommand{\aref}[1]{Appendix \ref{app:#1}}
\newcommand{\ket}[1]{|#1\rangle}
\newcommand{\bra}[1]{\langle#1|}
\newcommand{\ip}[2]{\langle#1|#2\rangle}

\newcommand{\Lop}{\hat{\cal L}}
\newcommand{\g}{\hat{g}}
\newcommand{\rhat}{\hat{r}}
\newcommand{\q}{\hat{q}}
\newcommand{\s}{\hat{s}}
\newcommand{\gam}{\hat{\gamma}}
\newcommand{\Gam}{\hat{\Gamma}}
\newcommand{\lam}{\hat{\lambda}}
\newcommand{\Lam}{\hat{\Lambda}}
\renewcommand{\ap}{\hat{a}^+}
\newcommand{\am}{\hat{a}^-}
\newcommand{\bp}{\hat{b}^+}
\newcommand{\bm}{\hat{b}^-}
\newcommand{\gb}{\bar{g}}
\newcommand{\qb}{\bar{q}}

\newcommand{\rb}{\bar{r}}

\begin{document}

\title{Noise expands the response range of the {\it Bacillus subtilis} competence circuit}

\author{Andrew Mugler\footnote{These authors contributed equally to this work.}}
\affiliation{Department of Physics, Purdue University, West Lafayette, IN 47907, USA}

\author{Mark Kittisopikul\footnotemark[1]}
\affiliation{Department of Biophysics, University of Texas Southwestern Medical Center, Dallas, TX 75390, USA}
\affiliation{Division of Biological Sciences, University of California San Diego, CA 92093, USA}

\author{Luke Hayden}
\affiliation{Division of Natural Sciences, Indiana Wesleyan University, Marion, IN 46953, USA}

\author{Jintao Liu}
\affiliation{Division of Biological Sciences, University of California San Diego, CA 92093, USA}

\author{Chris H.\ Wiggins}
\affiliation{Department of Applied Physics and Applied Mathematics, Columbia University, New York, NY 10027, USA}

\author{G{\"u}rol M.\ S{\"u}el}
\email{gsuel@ucsd.edu}
\affiliation{Division of Biological Sciences, University of California San Diego, CA 92093, USA}

\author{Aleksandra M.\ Walczak}
\email{awalczak@lpt.ens.fr}
\affiliation{Laboratoire de Physique Th\'eorique, CNRS, Universit\'e Pierre et Marie Curie and \'Ecole Normale Sup\'erieure, 75005 Paris, France}

\begin{abstract}
Gene regulatory circuits must contend with intrinsic noise that arises due to finite numbers of proteins.  While some circuits act to reduce this noise, others appear to exploit it.  A striking example is the competence circuit in {\em Bacillus subtilis}, which exhibits much larger noise in the duration of its competence events than a synthetically constructed analog that performs the same function.  Here, using stochastic modeling and fluorescence microscopy, we show that this larger noise allows cells to exit terminal phenotypic states, which expands the range of stress levels to which cells are responsive and leads to phenotypic heterogeneity at the population level.  This is an important example of how noise confers a functional benefit in a genetic decision-making circuit.
\end{abstract}

\maketitle

\section{Author Summary}
Fluctuations, or ``noise'', in the response of a system is usually thought to be harmful. However, it is becoming increasingly clear that in single-celled organisms, noise can sometimes help cells survive. This is because noise can enhance the diversity of responses within a cell population. In this study, we identify a novel benefit of noise in the competence response of a population of {\it Bacillus subtilis} bacteria, where competence is the ability of bacteria to take in DNA from their environment when under stress. We use computational modeling and experiments to show that noise increases the range of stress levels for which these bacteria exhibit a highly dynamic response, meaning that they are neither unresponsive, nor permanently in the competent state. Since a dynamic response is thought to be optimal for survival, this study suggests that noise is exploited to increase the fitness of the bacterial population.

\begin{figure*}
\centering
\includegraphics[width=\textwidth]{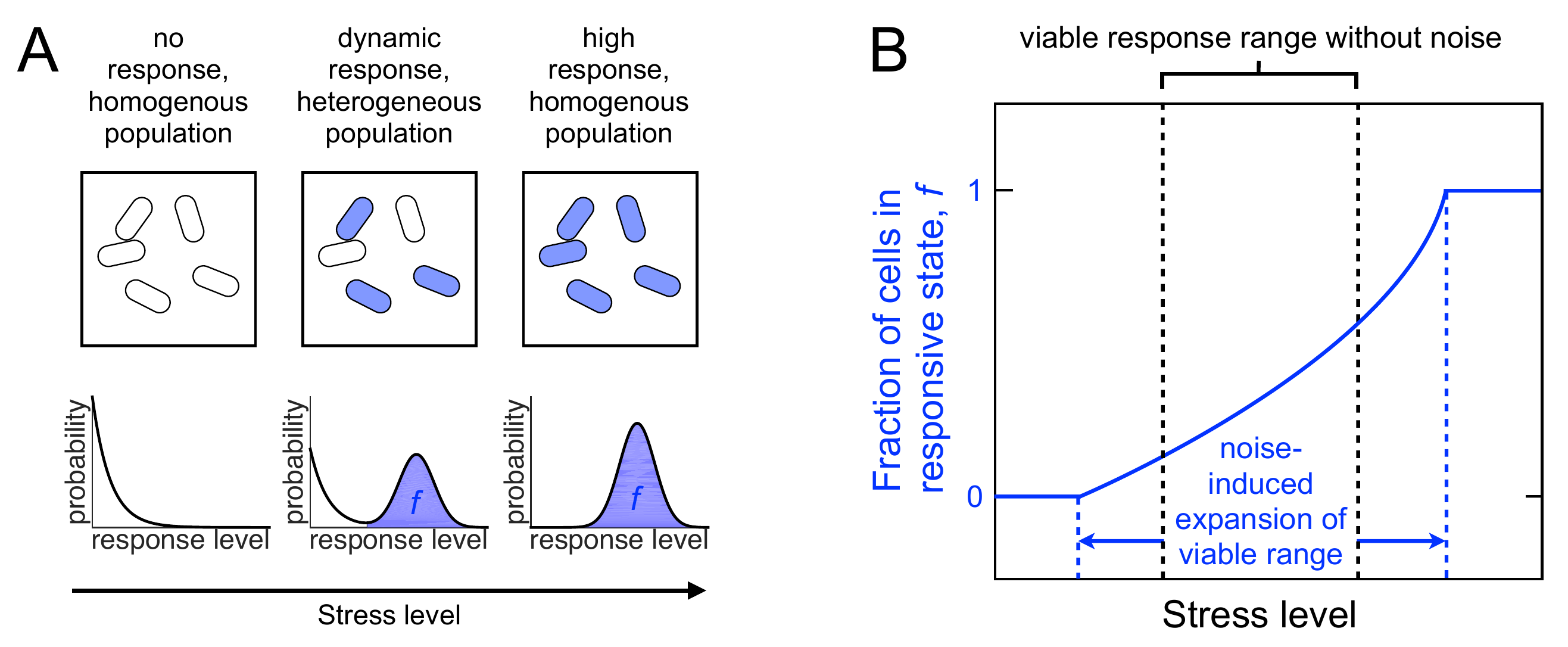}
\caption{{\bf Schematic illustrating phenotypic heterogeneity and the effects of noise.} (A) When all cells in a population exhibit either no response (left) or a high response (right), then the population is homogenous. In contrast, if individual cells exhibit a dynamic response (middle), this leads to a heterogenous population, with a fraction $f$ of cells in the responsive state at any given time. (B) Intrinsic noise affects the dynamics of the response. For the {\it B.\ subtilis} competence response, we find in this study that noise expands the viable response range: the range of stress levels over which $f$ remains neither $0$ nor $1$.}
\flabel{cartoon}
\end{figure*}

\section{Introduction}
Snapshots of bacterial populations often reveal large phenotypic heterogeneity in the gene expression states of its composite individuals. Such phenotypic heterogeneity in a clonal population of bacterial cells in a single environment has significant consequences for how well the organisms can adapt and survive. On the one hand, a population with little or no heterogeneity may allow for all cells to take advantage of certain optimal conditions to which the population is exposed. In this case, heterogeneity is suboptimal and therefore detrimental to fitness. On the other hand, numerous recent studies have shown that heterogeneous populations allow for cells to account for uncertainty in future environmental conditions \cite{fraser2009, kussell2005, balaban2004, kuchina2011a, kuchina2011b, mora2013}. In this case, heterogeneity is beneficial to fitness. A straightforward way to maintain high phenotypic heterogeneity is for each cell to exhibit a dynamic response. This allows each cell in its turn to transition among the various states of the population, e.g.\ via switching, pulsing, or oscillatory dynamics. The heterogeneity is intrinsically encoded in each cell, and is often enhanced by, or even entirely due to, stochasticity, or ``noise'', at the molecular level \cite{kussell2005, balaban2004, kuchina2011a, mora2013}.

The ability of molecular noise to cause stochastic phenotype changes has been demonstrated in a number of biological systems. In the context of enzymes, several studies have explored how intrinsic noise due to low numbers of molecules, or even a single molecule, can have dramatic effects through the amplified actions of a few enzymes \cite{novick1957, choi2008}. Moreover, studies of bacterial operons, including in the context of bacterial persistence, have suggested that stochasticity could be encoded in the interactions between genes in a genetic regulatory network by ensuring that certain operon states are exposed to low numbers of molecules \cite{savageau1974, kittisopikul2010, koh2012}. Recently, a theoretical study has demonstrated the conditions for when deterministic approaches to modeling genetic circuit dynamics break down, due to amplified effects of rare events caused by a small number of regulators \cite{michel2013}. Together, these works suggest that phenotypic heterogeneity could be rooted in low-molecule-number noise, and that this noise could in turn be encoded in the architecture of genetic regulatory networks.

The competence response of the gram-positive bacterium {\it Bacillus subtilis} provides a striking example of dynamically maintained phenotypic heterogeneity. Under stress, {\it B.\ subtilis} undergoes a natural and transient differentiation event, termed competence, that allows the organism to incorporate exogenous genes into its genome. Previous studies have shown that entry into the competent state is controlled by a genetic circuit that that can be tuned to one of three dynamical regimes \cite{suel2007}: an excitable regime at low stress levels, where cells rarely and transiently enter the competent state; an oscillatory regime at intermediate stress, where cells oscillate in and out of the competent state; and a mono-stable regime at high stress, where cells remain in the competent state. Importantly, oscillatory (and repeatably excitable) dynamics lead to phenotypic heterogeneity, since cells are dynamically transitioning in and out of the competent state (see \fref{cartoon}A). This heterogeneity is especially important to the survival of {\it B.\ subtilis}: if no cells respond, competence is not exploited, and the population may succumb to the stress. On the other hand, if all cells are permanently in the competent state, this can also be fatal to the population, since competence has been shown to reduce the cell growth rate and prevent cell division due to the inhibition of FtsZ \cite{haijema2001, suel2006}. Therefore, maintaining a dynamic competence response, and therefore a heterogenous population, is thought to be crucial to survival under stress.

\begin{figure*}
\centering
\includegraphics[width=\textwidth]{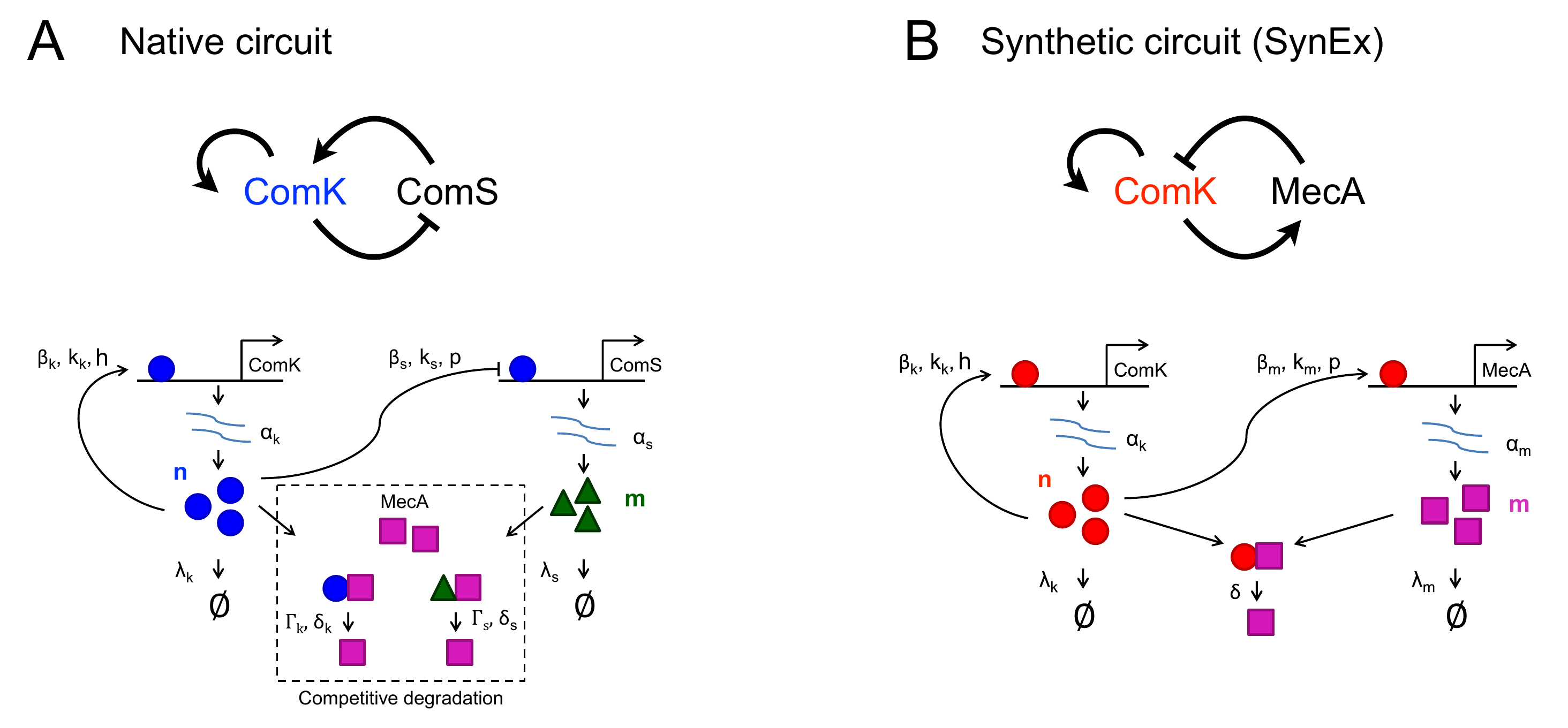}
\caption{{\bf Architectures and model parameters of the native and SynEx circuits.} The top row summarizes the regulatory interactions, while the bottom row depicts the model details. (A) In the native circuit, ComK is produced with the induction rate $\alpha_k$ and activates its own expression with Hill function parameters $\beta_k$, $k_k$, and $h$. ComS is expressed at the basal rate $\alpha_s$ and is repressed by ComK with Hill function parameters $\beta_s$, $k_s$, and $p$. ComK and ComS are degraded at rates $\lambda_k$ and $\lambda_s$, respectively, and, additionally, both compete for binding to the degradation enzyme MecA. MecA degrades ComK and ComS with maximal rates $\delta_k$ and $\delta_s$, respectively, and with Michaelis-Menten constants $\Gamma_k$ and $\Gamma_s$, respectively. (B) In the SynEx circuit, ComK is produced with the induction rate $\alpha_k$ and activates its own expression with Hill function parameters $\beta_k$, $k_k$, and $h$. MecA is expressed at the basal rate $\alpha_m$ and is activated by ComK with Hill function parameters $\beta_m$, $k_m$, and $p$. ComK and MecA are degraded at rates $\lambda_k$ and $\lambda_m$, respectively, and MecA enzymatically degrades ComK with rate $\delta$.}
\flabel{diagram}
\end{figure*}

The effects of noise on the dynamics of the competence response are only partially understood. Previous work has shown that noise can trigger excitations into the competent state when the circuit is tuned to the excitable regime \cite{suel2006}. Later work showed further that these excitations have a high variability in their duration, and that this variability is directly linked to the architecture of the competence circuit \cite{cagatay2009}. In particular, this work employed an analogous synthetic excitable circuit, termed SynEx, to provide evidence that the duration variability is due to intrinsic noise from low molecule numbers in the native circuit. However, the ability of this intrinsic noise to trigger sustained or repeatable excitations has not yet been quantified. Moreover, the generic effects of intrinsic noise on the three dynamic regimes, and how these effects translate to the physiological function of {\it B.\ subtilis} at the population level, are unknown.

Here, using stochastic modeling and quantitative fluorescence microscopy, we study the effects of intrinsic noise on the competence dynamics and the ensuing population heterogeneity of {\it B.\ subtilis}. We uncover a novel effect of noise that goes beyond architecture-dependent stochastic effects in a single cell.
Specifically, we find that at both low and high stress levels, noise prevents cells from becoming unresponsive or indefinitely responsive to the stress, and instead allows cells to respond dynamically.
These effects expand the range of stress levels over which the population of cells maintains a heterogeneous response distribution, which is critical to the population viability (see \fref{cartoon}B). The use of efficient numerical methods and stochastic simulation at several levels of model complexity allows us to elucidate the mechanisms behind these effects. A central prediction from our modeling is that these effects are rooted in noise arising from low numbers of molecules.
We verify this prediction using quantitative fluorescence microscopy by comparing the population response of native {\it B.\ subtilis} with that of synthetic mutants harboring the less-noisy SynEx circuit. Taken together, these results constitute a fundamental example of how noise can increase the functionality of a phenotypic response.

\begin{figure*}
\centering
\includegraphics[width=\textwidth]{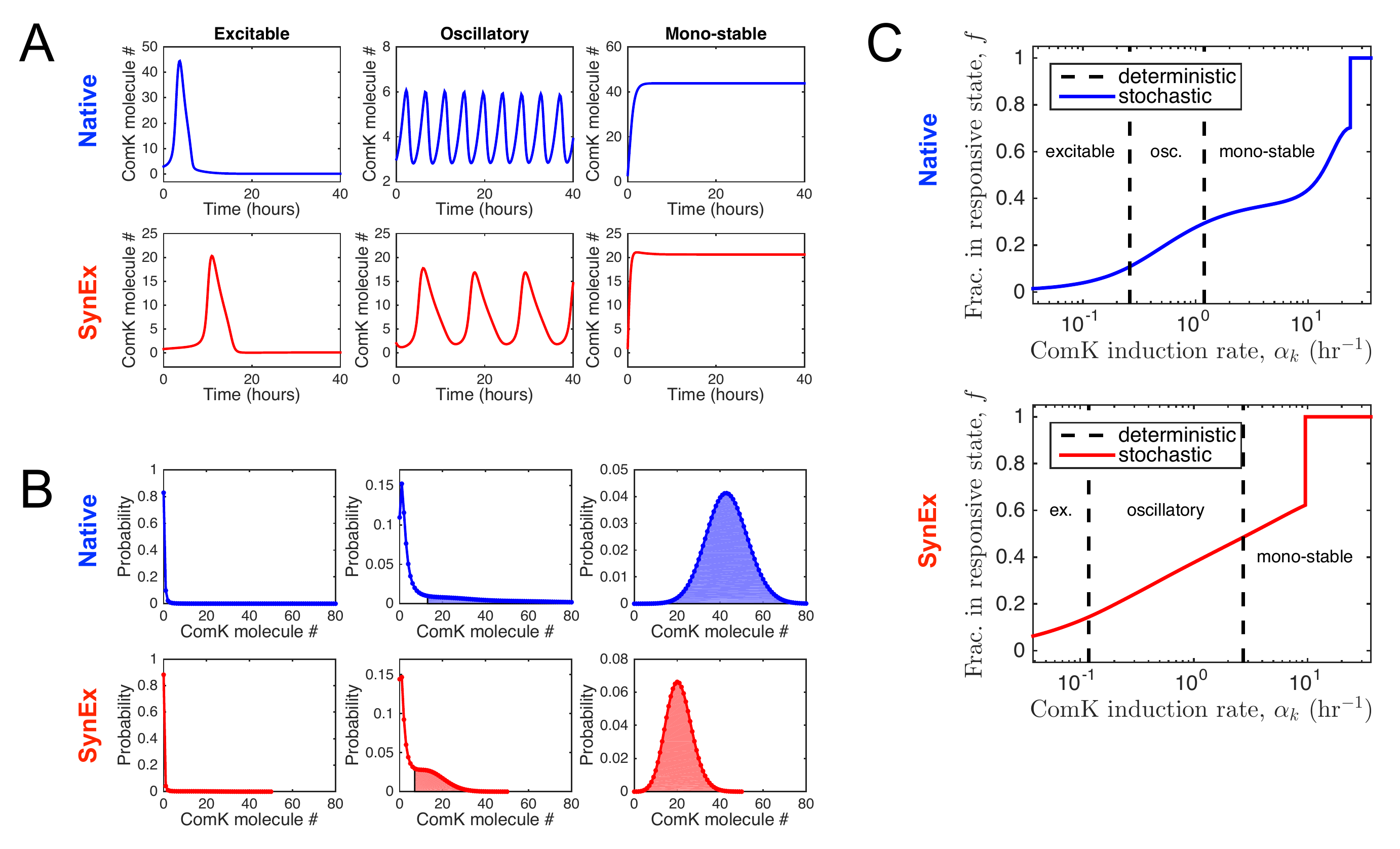}
\caption{{\bf Stochastic modeling of {\it B.\ subtilis} competence.} (A) The deterministic model of each circuit exhibits three dynamic regimes (excitable, oscillatory, and mono-stable), depending on the ComK induction rate $\alpha_k$, which models stress level. (B) The stochastic model reveals the ensuing distribution of response levels in each of the dynamic regimes. The fraction of the distribution in the responsive state $f$ (determined by the inflection points, see \sref{methods}) is shaded. (C) Whereas the deterministic model exhibits sharp transitions between the dynamic regimes (dashed lines), the stochastic model exhibits a continuous dependence of $f$ on induction rate. We see that for both circuits, stochasticity extends the viable response range ($0<f<1$) beyond the transitions predicted by the deterministic model, in both directions. Parameters are as in \cite{cagatay2009} and are given in \aref{SI}. In A and B, from left to right, the values of the control parameter are $\alpha_k = \{0.072, 1.15, 36\}$/hour (native) and $\alpha_k = \{0.036, 1.8, 36\}$/hour (SynEx).}
\flabel{theory}
\end{figure*}

\section{Results}

Entry of {\it B.\ subtilis} cells into the competent state occurs at high expression levels of the ComK protein. This protein activates a set of downstream genes allowing for the uptake of DNA \cite{suel2006}. ComK is typically expressed at a basal level, and stress in the environment alters the level of expression. In our genetic circuit design, as described below, increasing the stress level is mimicked by inducing comK expression using an increasing amount of a lactose analogue, Isopropyl $\beta$-D-1-thiogalactopyranoside (IPTG), in the environment.

In the native competence circuit, ComK activates its own expression, and represses the expression of another protein, ComS. ComS and ComK compete to be rapidly degraded by the MecA protein complex \cite{suel2006} (see \fref{diagram}A, bottom). Therefore, high concentrations of ComS hinder the degradation of ComK, effectively providing positive feedback to ComK by allowing ComK levels to build up. These interactions are summarized in \fref{diagram}A (top).

In the SynEx circuit, as described in \cite{cagatay2009}, the repression of ComS by ComK is removed by gene knockout. Then, the expression of MecA is placed under the control of ComK. This causes ComK to activate MecA, which in turn represses ComK via active protein degradation (see \fref{diagram}B, bottom). These interactions are summarized in \fref{diagram}B (top). Note that in the native circuit, ComK represses its own activator (ComS), while in the SynEx circuit, ComK activates its own repressor (MecA).

Both the native and SynEx circuits have architectures characteristic of molecular oscillators. Therefore we expect both circuits to allow for a dynamic response of each individual in a population. However, the main difference is that in the native circuit, when ComK levels are high, ComS levels are low, which leads to large amounts of intrinsic noise. In contrast, in the SynEx circuit, when ComK levels are high, MecA levels are also high, corresponding to less intrinsic noise. Previous work showed that this difference in architecture causes the native circuit to display a broad range of competence durations, whereas the SynEx circuit displays a relatively narrow range of competence durations \cite{cagatay2009}. However, the effects of noise and architecture on the ranges of dynamic response and the ensuing population heterogeneity in these systems remained unknown.

\begin{figure*}
\centering
\includegraphics[width=\textwidth]{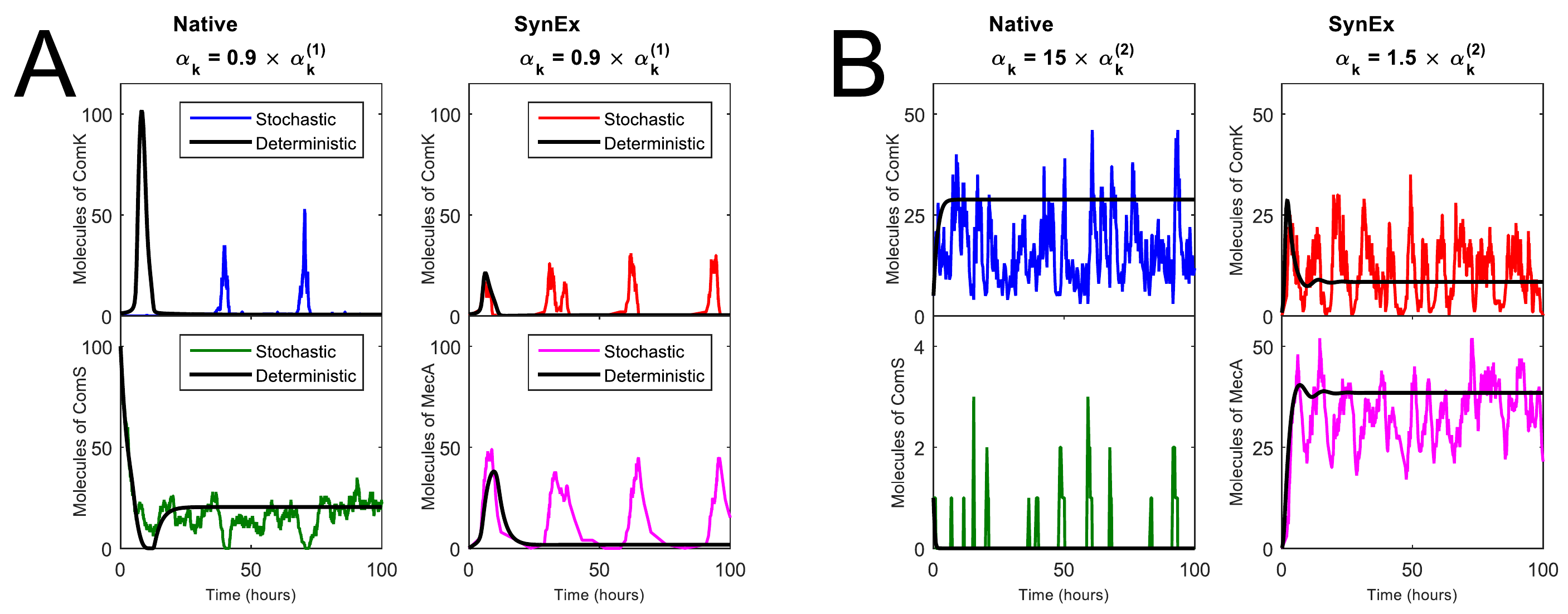}
\caption{{\bf Stochastic oscillations persist outside the deterministic oscillatory regime.} The deterministic oscillatory regime is defined by $\alpha_k^{(1)} < \alpha_k < \alpha_k^{(2)}$ for the induction rate $\alpha_k$. (A) At low induction rate $\alpha_k < \alpha_k^{(1)}$, where the deterministic model predicts excitable dynamics, the stochastic dynamics are oscillatory. The oscillations arise from repeated noise-induced excitations. (B) At high induction rate $\alpha_k > \alpha_k^{(2)}$, where the deterministic model predicts mono-stable dynamics, the stochastic dynamics are also oscillatory. The oscillations here arise because noise prevents damping to the mono-stable state (see the deterministic curves in the right panels). The effect is much stronger for the native circuit (notice that the left panel is 15 times outside the deterministically oscillatory regime) because, unlike in the SynEx circuit, one of the species, ComS, is at low copy number and therefore subject to significant intrinsic noise.}
\flabel{sim}
\end{figure*}

\subsection{Noise expands the response range}

To elucidate the effects of noise in each of the native and SynEx circuits (\fref{diagram}), we develop a stochastic model of each circuit, which includes noise, and then compare each to its deterministic analog, which does not include noise. As described in \sref{methods}, {\it Materials and Methods}, we develop the stochastic models at several levels of complexity to investigate the robustness of our findings to our modeling assumptions, and we solve each model using a combination of efficient numerical solution and stochastic simulation.  We first describe the behavior of the deterministic models. As shown in \fref{theory}A, a standard linear stability analysis of the deterministic model for each circuit reveals three dynamical regimes, depending on the value of the control parameter, the ComK induction rate $\alpha_k$. At low induction, each circuit is excitable, resulting in a transient differentiation event into and out of the competent (high-ComK) state. At intermediate induction, each circuit is oscillatory, periodically entering and exiting the competent state. At high induction, each circuit is mono-stable, staying in the competent state indefinitely. These three dynamical regimes have been confirmed in experimental studies of the native competence circuit \cite{suel2007}.

We find that these deterministic dynamics are reflected in the stationary solutions to the minimal stochastic models. As shown in \fref{theory}B, the three types of dynamics correspond to three shapes of stationary probability distributions of ComK levels. Excitable dynamics correspond to a distribution confined to low ComK molecule numbers, oscillatory dynamics correspond to a distribution mixed between low and high molecule numbers, and mono-stable dynamics correspond to a distribution centered at high molecule numbers. As described in \sref{methods}, we calculate the fraction $f$ of the distribution in the high-molecule-number state (see the shaded regions in \fref{theory}B). Within our model, $f$ represents the fraction of time a single cell spends in the competent state, or equivalently, the fraction of an isogenic population of cells found in the competent state at a given time. Importantly, $f$ is the indicator of population heterogeneity, since unresponsive ($f=0$) or fully competent ($f=1$) populations are homogeneous, while mixed populations ($0<f<1$) are heterogeneous. We define the range of induction rate $\alpha_k$ for which $0<f<1$ as the {\it viable response range}, since unresponsive cells ($f=0$) do not benefit from competence, while long-term competence ($f=1$) is known to have a detrimental effect on growth rate and cell division \cite{haijema2001, suel2006}.

In \fref{theory}C, we compare the viable response range of the stochastic model with the boundaries between dynamical regimes predicted by the deterministic model. We see that for both the native and the SynEx circuit, the stochastic range extends beyond the deterministic range for both low and high induction rate $\alpha_k$. Furthermore, the extension at high induction rate is significantly more pronounced for the native circuit (roughly $20$ times the deterministic value) than for the SynEx circuit (roughly $3$ times the deterministic value). These observations imply that noise expands the range of stress levels to which cells can respond in a dynamic way. In the next section, we elucidate the mechanisms behind this expansion.

\subsection{Noise-induced oscillations underlie the expansion of the response range}

Why does noise expand the viable response range at low induction levels? As shown in \fref{sim}A, the reason is that noise leads to repeated excitations into the competent state, which prevents the system from remaining completely unresponsive. In a completely deterministic excitable system, an excitation is caused by initializing the system away from its stable fixed point, and it occurs only once. However, in a stochastic system, noise can cause repeated perturbations away from the stable state, leading to persistent additional excitations. Indeed, in both circuits, noise at the stable state is high, because the stable state corresponds to one or more species being expressed at very low molecule number (ComK for the native circuit, ComK and MecA for the SynEx circuit; see \fref{sim}A). Since the dynamics are governed by  Poissonian birth-death reactions, low molecule numbers correspond to high intrinsic noise (variance over the squared mean), leading to frequent and persistent excitations. This effect is consistent with the noise-induced excitations seen for these circuits in previous work \cite{suel2006,cagatay2009}. Here, however, we have quantified the effect of these excitations on the stochastic distribution, which describes the heterogeneous population response.

Why does noise expand the viable response range at high induction levels? Here the mechanism is different from at low induction levels. As shown in \fref{sim}B, the reason is that noise prevents the damping of oscillations, which keeps the system from relaxing to the competent state. In the deterministic system, the mono-stable state is defined by a stability matrix whose eigenvalues are complex with negative real parts (\fref{stability}). This means that the solution relaxes to the mono-stable state in an oscillatory way, i.e.\ the oscillations are damped (see the black lines in the right panel of \fref{sim}B, for example). Intrinsic noise thwarts this relaxation, continually perturbing the system away from the stable point, and preserving a finite oscillation amplitude (see the colored lines). Similar effects have been observed in ecological and epidemic models, where they are attributed to the ability of white noise to repeatedly excite a system at its resonant frequency \cite{black2012}. Here we see the effect at the molecular level in bacteria, and we find that it occurs sufficiently strongly that it supports and significantly extends a heterogeneous population response.

At high induction levels, the expansion of the viable response range is more pronounced in the native circuit than in the SynEx circuit. This effect was demonstrated at the population level in \fref{theory}C. It is also demonstrated by the dynamics in \fref{sim}B: the noise-induced prevention of damping is clearly evident for the native circuit, even at the $\alpha_k$ value shown, which is $15$ times value predicted deterministically. The reason that the effect is so pronounced in the native circuit is that the mono-stable fixed point corresponds to ComS being expressed at very low molecule numbers, where the intrinsic noise is high (lower left panel). In contrast, in the SynEx circuit, the mono-stable state corresponds to both species begin expressed at higher molecule numbers, so the intrinsic noise is lower. This difference, which stems ultimately from the difference in the architecture of the two circuits (\fref{diagram}), was found in previous work \cite{cagatay2009} to be responsible for the increased variability in the competence durations of the native circuit compared to the SynEx circuit. Here we demonstrate that the architecture of the native circuit additionally leads to an increase in the expansion of its viable response range, which has a clear benefit for fitness.

We have tested that the effects discussed above are robust, in that they persist when we relax the three simplifying assumptions of our minimal stochastic model (see \sref{methods}). We relax two of the assumptions by considering a non-adiabatic stochastic model in which the fast dynamics of mRNA production and enzymatic degradation are included explicitly, and by setting the mean molecule numbers in the tens of thousands as opposed to tens (see \aref{SI}). We find that all noise-induced effects persist, namely (i) repeated excitations, (ii) the prevention of damping, and (iii) the enhancement of effect ii in the native circuit over the SynEx circuit (see \fref{native_full} and \fref{synex_full}). As shown in \fref{native_full} and \fref{synex_full}, we also verified quantitatively that effects i and ii produce sufficiently oscillatory dynamics that the power spectrum is peaked, as opposed to the non-peaked power spectrum observed for purely excitable or mono-stable dynamics. Interestingly, when we raise the molecule number, but retain the adiabatic assumption, we find that the effects of noise diminish, and the stochastic model behaves like the deterministic model (see \fref{native_high} and \fref{synex_high}). This confirms that the effects we observe are rooted in the intrinsic noise arising from low molecule numbers, as expected. Importantly, however, it demonstrates that when coupled with explicit mRNA and competitive degradation dynamics, these intrinsic effects dominate the response up to a much higher molecule number regime.

Finally, we relax the third assumption by considering a three-species model for the SynExSlow circuit, in which the dynamics of ComS are accounted for explicitly (see \fref{synexslow}). We find that effect ii persists, while effect i does not, indicating that the expansion of the viable response regime at high induction levels is more robust than at low induction levels. Since this is also the more pronounced effect, we focus on the high-induction regime in the next section, where we compare our model predictions with experiments.

\subsection{Fluorescence microscopy confirms the predictions of the model}

To test our model predictions, we use quantitative fluorescence microscopy to measure the ComK expression levels in populations of {\it B.\ subtilis} cells harboring either the native or the SynExSlow circuits, as described in \sref{methods} (see \fref{expt}A). ComK expression is induced by increasing the concentration of IPTG, which corresponds to the model parameter $\alpha_k$. As seen in \fref{expt}B, in both the native and the SynEx strain, as the IPTG concentration increases, the fluorescence distribution across the population changes shape: first it is centered at low values, then it is split between low and high values, and finally it is centered at high values. This change is qualitatively reminiscent of the change seen in the stochastic model in \fref{theory}B. Moreover, \fref{expt}C also shows that the transition to a distribution centered at high values occurs at a higher IPTG concentration in cells with the native circuit than in cells with the SynExSlow circuit.  This feature is also qualitatively consistent with the theoretical prediction shown in \fref{theory}C.

\begin{figure*}
\centering
\includegraphics[width=\textwidth]{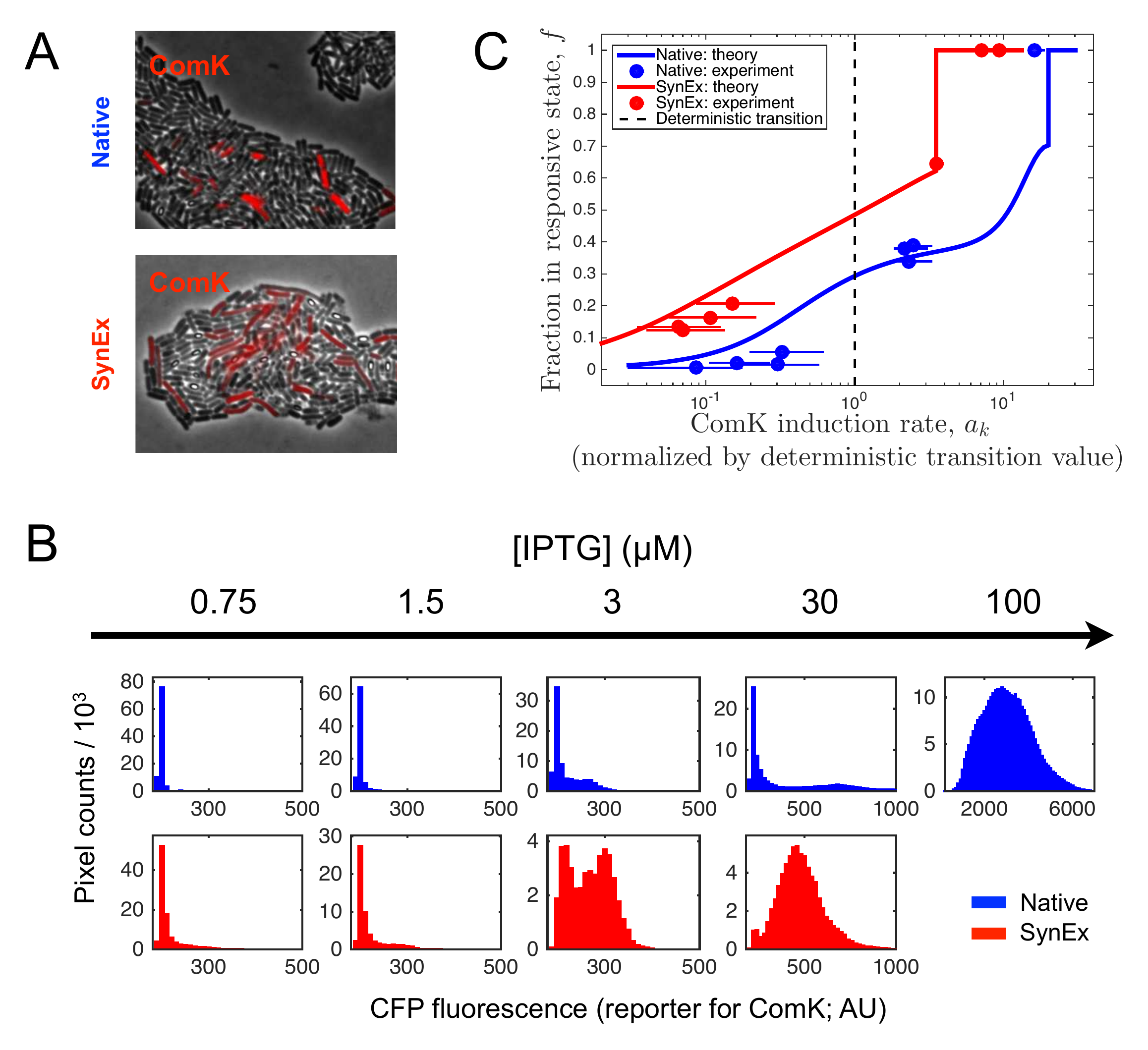}
\caption{{\bf Quantitative fluorescence microscopy confirms model predictions.} (A) Microscopy image of {\it B.\ subtilis} cells with 1.5 $\mu$M IPTG. ComK expression, measured by CFP fluorescence, is shown in red. (B) Fluorescence distributions over the imaged populations for both circuits as a function of IPTG levels. Note that, as in the model, the shift from a non-responsive state (low fluorescence) to a responsive state (high fluorescence) is clearly evident in the distributions. (C) Comparison of the data with the model in the high-induction regime. For both circuits, $\alpha_k$ is normalized by the value of the deterministic transition from the oscillatory to the mono-stable regime (dashed line). Agreement between the model and data confirms both model predictions: that noise extends the viable response regime to higher stress levels than predicted deterministically, and that the effect is more pronounced in the native circuit than in the SynEx circuit.}
\flabel{expt}
\end{figure*}

To investigate whether our experimental observations agree quantitatively with our theoretical predictions, as well as qualitatively, we fit the fluorescence distributions to the stochastic model, as described in \sref{methods}.  \fref{expt}C shows that both of our central predictions in the high-induction regime are quantitatively confirmed by the data, namely (i) that noise extends the transition to a permanently competent state beyond the deterministically predicted induction level, and (ii) that it does so to a larger extent in the native circuit than in the SynEx circuit. \fref{expt} therefore provides strong experimental support for the the notion that intrinsic noise expands the viable response range by delaying, as a function of induction level, the relaxation of cells to the competent state.

\section{Discussion}

Phenotypic heterogeneity, in which different individuals express particular genes at different levels, is an important survival strategy in uncertain environments.
Here we studied dynamically maintained phenotypic heterogeneity in the competence response of {\it B.\ subtilis}, and how it is influenced by intrinsic fluctuations in molecule numbers. By combining theoretical modeling, stochastic simulations, and quantitative microscopy, we showed that intrinsic noise facilitates heterogeneity by expanding the range of stress levels over which heterogeneity is maintained (the viable response range). The effect manifests itself at both low and high stress levels, and the influence of noise is dramatic: in the native competence circuit, noise increases the maximal stress level at which a heterogeneous population response occurs, by $20$-fold.

Our work advances previous work investigating the effects of circuit architecture on dynamic response. It was previously known that the native competence circuit exhibited higher variability in its competence duration times than a synthetic analog with different architecture (SynEx). This variability was attributed to intrinsic molecule number fluctuations and was thought to provide a fitness advantage, similar to variability in the times to commit to cell states \cite{nachman2007}. Yet the advantages of the native design over the synthetic design were not immediately clear. Here, we showed that while the SynEx circuit is more predictable in terms of competence duration times, its dynamic response is limited to a much smaller range of stress levels, which limits its functionality.

In both circuits, the viable response range is expanded at both low and high stress levels. The mechanisms in these two cases are different. At low stress levels, intrinsic noise causes repeated, period excitations, effectively sustaining oscillations into the excitable regime. At high stress levels, noise prevents the damping of oscillations, effectively delaying, as a function of stress level, the static and indefinite entry into the competent state. Both mechanisms rely on intrinsic fluctuations and, importantly, persist even at high molecule numbers in a non-adiabatic system. The limited response range observed in the deterministic solution is recovered only in the strict limit of fast switching and high molecule numbers, suggesting that the effects of noise that we observe here are generic.

Quantitative microscopy measurements confirmed our theoretical predictions:  all cells exhibited competence at high induction levels, no cells exhibited competence at low induction levels, and a bimodal population response was observed in the intermediate regime. Important differences between circuit architectures were also confirmed experimentally, namely that the SynEx circuit begins to oscillate at lower stress levels (IPTG levels) than the native circuit, and that the native circuit can withstand roughly $5$-fold higher stress levels than the SynEx circuit before indefinitely entering the competent state. An independent fit of the model predictions to the experimental data showed very good agreement.

Traditionally, noise in gene expression has often been seen as a nuisance that needs to be controlled, especially in stable environments or when the reproducibility of downstream gene expression is crucial. Thus, much work has concentrated on how to ensure the reliability of gene expression and cell signaling in the presence of intrinsic noise \cite{mugler2013, govern2014a, govern2014b, tostevin2007, saunders2009, tkacik2009, walczak2010, tkacik2012, tkacik2011}. However, as has been shown experimentally and theoretically in the context of antibiotic resistance \cite{balaban2004, rotem2010, gefen2009}, noise-induced population heterogeneity can be advantageous for adaptation to new conditions \cite{sato2003, ito2009, kashiwagi2006, shimizu2011}. Functional applications of noise have also been identified in a number of settings \cite{eldar2010} ranging from differentiation decisions to sporulate \cite{veening2008}, apoptose \cite{spencer2009}, or allow DNA uptake, such as discussed in this paper.

In {\it B.\ subtilis}, a heterogeneous competence response is thought to be optimal since permanent competence curbs cell growth. The effect of phenotypic heterogeneity on the growth rate of populations has also been studied theoretically \cite{thattai2004, kussell2005, kussell2005b, leibler2010, rivoire2011, sato2006, tanase2008}, showing that while in optimal conditions fluctuations decrease the overall growth rate, in less favorable environments, diversity of gene expression increases the population fitness \cite{tanase2008}. The effect of selection on such populations was also considered \cite{sato2006}, and shown to influence the stability of the phenotypic states \cite{mora2013}.

We have described an example of phenotypic heterogeneity that is maintained by an oscillatory response, and we have demonstrated that intrinsic noise increases the range of stress levels for which oscillations occur. The ability of noise to facilitate oscillations has also been observed in the entrainment of NF-$\kappa$B in fibroblast cells to oscillating TNF inputs \cite{kellogg2015}. There, small-molecule-number noise was shown to facilitate both oscillation and entrainment, and phenotypic variability was shown to enlarge the dynamic range of inputs for which entrainment is possible. These results, along with our findings herein, suggest that strategies that exploit the coupling between noise and phenotypic heterogeneity allow for functional population responses over a large variety of conditions.

\section{Materials and Methods}
\slabel{methods}

\subsection{Stochastic model}

Our minimal stochastic models of the native and SynEx circuits are based on our previous modeling work \cite{cagatay2009}, but employ the (stochastic) master equation instead of a (deterministic) dynamical system in order to capture the effects of intrinsic noise. The master equation describes the dynamics of the probability distribution over the numbers of the relevant molecular species inside the cell \cite{vankampen1992}. For both circuits the master equation reads
\beqn
\elabel{me}
\frac{dp_{nm}}{dt} &=& g_{n-1} p_{n-1,m} + r_{n+1,m} (n+1) p_{n+1,m} \nonumber\\
&&	+  q_n p_{n,m-1} + s_{n,m+1} (m+1) p_{n,m+1} \nonumber\\
&& - (g_n + r_{nm}n + q_n + s_{nm}m) p_{nm}.
\eeqn
where $p_{nm}$ is the joint probability distribution over molecule numbers $n$ and $m$ (see \fref{diagram} for a diagram and explanation of all variables and parameters). In the native circuit, $n$ is the number of ComK proteins and $m$ is the number of ComS proteins. In the SynEx circuit, $n$ is the number of ComK proteins and $m$ is the number of MecA proteins. The dynamics are birth-death processes with mutual regulation: the production rates $g$ and $q$ increase the numbers $n$ and $m$, respectively, while the degradation rates $r$ and $s$ decrease the numbers $n$ and $m$, respectively, and the regulation is encoded in the functional dependence of the rates on $n$ and $m$. The regulation functions follow from our previous work \cite{cagatay2009} and for the native circuit read
\begin{align}
\label{eq:r1}
g_n &= \alpha_k + \frac{\beta_k n^h}{k_k^h + n^h}, \\
\label{eq:r2}
q_n &= \alpha_s + \frac{\beta_s}{1 + (n/k_s)^p}, \\
\label{eq:r3}
r_{nm} &= \frac{\delta_k}{1 + n/\Gamma_k + m/\Gamma_s} + \lambda_k, \\
\label{eq:r4}
s_{nm} &= \frac{\delta_s}{1 + n/\Gamma_k + m/\Gamma_s} + \lambda_s,
\end{align}
while for the SynEx circuit they read
\begin{align}
\label{eq:r5}
g_n &= \alpha_k + \frac{\beta_k n^h}{k_k^h + n^h}, \\
\label{eq:r6}
q_n &= \alpha_m + \frac{\beta_m n^p}{k_m^p + n^p}, \\
\label{eq:r7}
r_{nm} &= r_m = \delta m + \lambda_k, \\
\label{eq:r8}
s_{nm} &= s = \lambda_m.
\end{align}
The meaning of the parameters is explained in \fref{diagram}. The regulatory functions introduce positive and negative feedbacks (see \fref{diagram}, top). The parameter values used in the model are as in \cite{cagatay2009} and are given in \aref{SI}.

The model in \erefn{me}{r8} makes three simplifying assumptions, all of which we later relax. First, as in \cite{suel2006, cagatay2009} we have assumed that mRNA dynamics and the enzymatic degradation process are substantially faster than all other biochemical reactions in the circuits, and are thus adiabatically eliminated (see \aref{SI} for details). This reduces each model to the two-species form in \eref{me}, depicted by the cartoons in \fref{diagram} (top). Second, the parameters are chosen such that typical protein copy numbers are small (in the tens or hundreds per cell). Lacking information about the absolute protein numbers in the experiments, we make this assumption because we expect any effects of intrinsic noise to be most evident in the low-number regime, although as we later show, the effects we find persist out to protein numbers in the tens of thousands. Third, we model in \erefn{r5}{r8} the SynEx circuit as originally constructed \cite{cagatay2009}, instead of the ``SynExSlow'' circuit that we use in experiments (described later in this section). This reduces the model from three species to two, which is more amenable to analytic and numerical solution. Once again, however, we will see that the most important effects of noise that we elucidate are also present in a model of the SynExSlow circuit.

We solve \eref{me} in steady state in one of two ways. At low copy numbers, we use the spectral method \cite{walczak2009, mugler2009}, a hybrid analytic-numerical technique that exploits the eigenfunctions of the birth-death process. Derivation of the spectral solution of \eref{me} is given in \aref{SI}. The spectral method is much more efficient than other numerical techniques \cite{walczak2009}, but we find here that it becomes numerically unstable at sufficiently high copy numbers. Therefore, at high copy numbers, we use iterative inversion of the matrix acting on $p_{nm}$ on the right-hand side of \eref{me} (see \aref{SI}). To obtain individual stochastic trajectories of the system described by \erefn{me}{r8}, we use the Gillespie algorithm \cite{gillespie1977}.

\subsection{Deterministic model}

The deterministic analog of \eref{me} is obtained by performing an expansion in the limit of large molecule numbers \cite{vankampen1992}. To first order one obtains
\beqn
\elabel{ode1}
\frac{d\bar{n}}{dt} &=& g_{\bar{n}} - r_{\bar{n}\bar{m}} \bar{n}, \\
\elabel{ode2}
\frac{d\bar{m}}{dt} &=& q_{\bar{n}} - s_{\bar{n}\bar{m}} \bar{m},
\eeqn
where $\bar{n}$ and $\bar{m}$ are ensemble averages. \erefs{ode1}{ode2} form a coupled dynamical system whose properties we obtain by linear stability analysis. As shown in \fref{stability}, both circuits exhibit excitable, oscillatory, and mono-stable regimes, depending on the value of the control parameter $\alpha_k$. The transition from excitable to oscillatory is marked by the annihilation of a stable and an unstable fixed point, leaving only one unstable fixed point. The transition from oscillatory to mono-stable is marked by this unstable fixed point becoming stable. These transitions provide the dashed lines in \fref{theory}C.

\subsection{Genetic circuit construction}

For the native competence circuit, we used a variant from our previous study \cite{suel2007}. For the synthetic competence circuit, we reconfigured the original ``SynExSlow'' circuit created in \cite{cagatay2009}, in order to introduce a tunable proxy for stress level. This required replacing the tunable $ P_{hyperspank}-comS $ with an internally controlled promoter for the ribosomal gene $ rpsD $, as well as adding in $ P_{hyperspank}-comK $. The result was a strain that is resistant to four antibiotics and has $ comS $ expressed from a ribosomal promoter, providing for a basal level of expression (see \aref{SI} for chromosomal alterations and antibiotic resistance). In both strains, ComK expression is induced by increasing the amount of IPTG in the environment. Since stress signals are usually integrated at the ComK promoter, IPTG therefore acts as a proxy for stress and triggers competence. This allows us to simulate stress directly in a controlled manner, rather than using physiological stresses that may themselves induce external variation in the responses.

\subsection{Time-lapse microscopy}

Cells of {\em Bacillus subtilis} were prepared by streaking from glycerol stocks onto LB agar plates containing the appropriate antibiotic for maintenance and incubated at 37 \textdegree C overnight.. Single colonies were then selected from the plates and grown in LB broth for three to four hours at 37 \textdegree C until an OD of 1.6 to 1.8 was reached. While culturing the cells, argarose pads were made by pouring 6 mL of 0.8\% w/v low-melting point agarose in re-suspension medium onto a glass coverslip. A second glass coverslip was then placed on top of the medium, and the medium was left to congeal while the culture was grown. Once the culture was ready, cells were spun down and resuspended in the resuspension medium twice to wash way the LB. To deposit cells, the top glass coverslip was removed, and then 2 $\mu$L of cells were dropped on 37 \textdegree C low melting point agarose pads. The pads were then cut into squares with a 5mm edge, each containing a single drop of cells. After drying for one additional hour, the pads were flipped over and placed on a glass-bottom dish. The dish was then sealed with parafilm. Images of the cells were then obtained at 100X magnification on an Olympus IX81 system using the ImagePro software from MediaCybernetics along with customized macros.

IPTG stock solutions were dissolved in ethanol to a concentration of 100 mM. Working (1000X) stocks were diluted with Milli-Q water to 30 mM, 10 mM, 3 mM, 1.5 mM, 0.75 mM, respectively, by serial dilution and then added to the appropriate media at a ratio of 1:999 to achieve the final concentrations indicated.

\subsection{Plasmid and strain construction}

Template plasmids with homologous recombination arms for the {\em Bacillus subtilis} chromosomal loci were modified through restriction enzyme digest and ligation of DNA inserts (see \aref{SI} for loci). The inserts were created by polymerase chain reactions using primers from Integrated DNA Technologies while using genomic DNA or other plasmids as templates.

The PY79 strain of {\em Bacillus subtilis} was modified through homologous recombination using a One-Step Transformation protocol by inducing competence. 50 ng of plasmid DNA was replicated in TOP10 {\em E. coli} cells (Invitrogen, Life Sciences, Inc) and purified using a MiniPrep spin column (Sigma-Aldrich). The DNA was then mixed with culture growing in minimal salts for thirty minutes and then subsequently were rescued using 2xYT rich medium. Positive colonies were then selected on LB agar plates containing selective concentrations of antibiotics.

\subsection{Image analysis}
Fluorescence histograms were obtained from microscopy images using a pixel-based
analysis. A mask was created on each image to identify
the areas that the cells occupy (see \fref{imaging}). A histogram
of fluorescence intensity values was then generated for pixels within that area.

\subsection{Culture media}

Sterlini-Mandelstram Resuspension Medium was used during time-lapse microscopy and followed the protocol as in references \cite{Sterlini1969,Harwood1990}. The actual protocol used consists of making two salt solutions: A and B. Solution A consists of 0.089 g of FeCl$_3\cdot 6$H$_2$O, 0.830 g of MgCl$_2\cdot 6$H$_2$O and 1.979 g MnCl$_2\cdot 4$H$_2$O in 100 mL of filtered water. Solution A is filter sterilized (not autoclaved) and stored at 4 \textdegree C. Solution B consists of 53.5 g NH$_4$Cl, 10.6 g Na$_2$SO$_4$, 6.8 g KH$_2$PO$_4$, and 9.7 g NH$_4$NO$_3$. Solution B is then also filter sterilized and stored at 4 \textdegree C. Sporulation salts are made by combining adding 1 mL of Solution A and 10 mL of Solution B to filtered water for a total 1 L. This solution is then autoclaved. The final Resuspension media is created by combining 93 mL of sporulation salts, 2 mL of 10\% v/v L-glutamate, 1 mL of 0.1M CaCl$_2$, and 4 mL of 1M MgSO$_4$ on the day of the experiment.

One-step transformation media consists of 6.25 g of K$_2$HPO$_4\cdot 3$H$_2$O, 1.5 g of KH$_2$PO$_4$, 0.25 g of trisodium citrate, 50 mg of MgSO$_4\cdot 7$H$_2$O, 0.5 g of Na$_2$SO$_4$ at pH 7.0, 125 $\mu$L of 100 mM FeCl$_3$, 5 $\mu$L of 100 mM MnSO$_4$, 1 g of glucose, and 0.5 g of glutamate added into filtered water for a total of 250 mL. The media is filter sterilized using 0.2 micron Millipore filters.

2xYT recovery medium consists of 16.0 g of Tryptone, 10.0 g of Yeast Extract, 5.0 g of NaCl added to filtered water to a total volume of 1L. The media is then filter sterilized using 0.2 micron Millipore filters.

\subsection{Distribution analysis and comparing experiments with modeling}

For the ComK distributions in the model, $p_n = \sum_m p_{nm}$, we determine the fraction $f$ of in the responsive, high comK protein concentration state using two independent methods. First, we use a generalized method of separating the distribution's two modes: since the distribution is often not completely bimodal (see \fref{theory}B, middle column), we find the average $n^* = (n_1 + n_2)/2$ of the two inflection points surrounding the putative local minimum between the two modes, and define $f = \sum_{n = n^*}^\infty p_n$. In the case of a bimodal $p_n$, this method indeed well approximates the location of the actual local minimum. Second, we fit $p_n$ to a mixture of two Poisson distributions, using the Kullback-Leibler divergence as the cost function. There are three fitting parameters, the two Poisson parameters and the relative weighting between them, and the  weighting provides $f$. We see in \fref{fcalc} that the two methods give similar results for the dependence of $f$ on the control parameter $\alpha_k$, demonstrating that our determination of $f$ is robust to the method used.

To compare the model predictions to the experimental data, we analyze the fluorescence distributions (\fref{expt}B) in the same  way as the model distributions. Specifically, we calculate the fraction of the comK population in the responsive state $f$ for each experimental distribution using the two-Poisson method  above. Because the mapping between IPTG concentration and the model parameter describing the level of external stress $\alpha_k$ is unknown, we infer the most likely value of $\alpha_k$ corresponding to each experimental distribution by fitting the theory to the data. First we use the mode of the $[$IPTG$] = 0$ distributions to subtract the background fluorescence from the remaining data. To avoid binning, we then fit the cumulative distribution instead of the probability distribution (sample fits are shown in \fref{fitting}). We use a maximally constrained least-squares fit, where all parameters are fixed as in \aref{SI} except $\alpha_k$ and the unknown parameter $X$ describing the conversion of pixel intensity to molecule number. Given a value of $X$, we find the values of $\alpha_k$ for each distribution that minimize the sum of the squared error $S$ in each case. $X$ is then chosen by minimizing the sum of minimum $S$ values over all distributions. These $f$ and $\alpha_k$ values inferred from the data are plotted in \fref{expt}C. The error bars on $\alpha_k$ are obtained by finding the $\alpha_k$ values where $S$ reaches $1.25$ of its minimum value (sample plots of $S$ vs.\ $\alpha_k$ are shown in \fref{fitting}).

\begin{acknowledgments}
M.K. was supported by NIH biophysics training grant T32GM008297 and acknowledges Tolga Cagatay for assistance in preparing Bacillus strains. L.H.\ was supported by NSF REU grant PHY-1460899. G.M.S. was supported by the National Institute of General Medical Sciences Grant R01 GM088428 and the National Science Foundation Grant MCB-1450867. A.M.W.\ was supported by MCCIG grant no. 303561. This work was also supported by the San Diego Center for Systems Biology (NIH Grant P50 GM085764).
\end{acknowledgments}

\onecolumngrid
\appendix

\section{Supporting Information}
\alabel{SI}

\subsection{Parameter values}
\slabel{param}

The parameter values used in the model in the main text are taken from \cite{suel2006, suel2007, cagatay2009}, where they have been optimized to agree with a number of experimental observations, including the existence of competence events, the duration of these events, and the robustness of events to a wide range of stress levels. They are first given below in dimensionless form, which is sufficient to establish their deterministic dynamical properties. To account for intrinsic noise and to compare with experiments, several physical quantities are then specified to establish molecule numbers and timescales. The conversion to the remaining model parameters, and their resulting values, are then given below.\\

Native circuit:\\

\begin{tabular}{|l|}
	\hline
	Dimensionless parameters \\
	\hline \hline
	$a_s=0$ \\
	$b_k=0.3$ \\
	$b_s=3$ \\
	$k_0=0.2$ \\
	$k_1=1/30$ \\
	$\Delta_k=0.1$ \\
	$\Delta_s=0.1$ \\
	$h=2$ \\
	$p=5$ \\
	\hline
\end{tabular}
\begin{tabular}{|l|}
	\hline
	Physical quantities \\
	\hline \hline
	Molecule numbers: \\
	$\Gamma_k = 100$ \\
	$\Gamma_s = 1$ \\
	\\
	Timescales: \\
	$\delta_k = 0.001$/s\\
	$\delta_s = 0.001$/s\\
	\\
	\\
	\hline
\end{tabular}
\begin{tabular}{|l|}
	\hline
	Remaining parameters \\
	\hline \hline
	$\alpha_k$ (varied) \\
	$\alpha_s = a_s\Gamma_s\delta = 0$/s \\
	$\beta_k = b_k\Gamma_k\delta_k = 0.03$/s \\
	$\beta_s = b_s\Gamma_s\delta_s = 0.003$/s \\
	$k_k = k_0\Gamma_k = 20$ \\
	$k_s = k_1\Gamma_k = 3.3$ \\
	$\lambda_k = \Delta_k\delta_k = 10^{-4}$/s \\
	$\lambda_s = \Delta_s\delta_s = 10^{-4}$/s \\
	\\
	\hline
\end{tabular}\\ \\

SynEx circuit:\\

\begin{tabular}{|l|}
	\hline
	Dimensionless parameters \\
	\hline \hline
	$a_m=0.3$ \\
	$b_k=15$ \\
	$b_m=10$ \\
	$\mu=1$ \\
	$h=2$ \\
	$p=2$ \\
	\\
	\hline
\end{tabular}
\begin{tabular}{|l|}
	\hline
	Physical quantities \\
	\hline \hline
	Molecule numbers: \\
	$k_k = 10$ \\
	$k_m = 5$ \\
	\\
	Timescales: \\
	$\lambda_k = 10^{-4}$/s\\
	$\lambda_m = 10^{-4}$/s\\
	\hline
\end{tabular}
\begin{tabular}{|l|}
	\hline
	Remaining parameters \\
	\hline \hline
	$\alpha_k$ (varied) \\
	$\alpha_m = a_m k_m\lambda_m = 1.5 \times 10^{-4}$/s \\
	$\beta_k = b_k k_k\lambda_k = 0.015$/s \\
	$\beta_m = b_m k_m \lambda_m = 0.005$/s \\
	$\delta = \mu\lambda_k/k_m = 2\times 10^{-5}$/s \\
	\\
	\\
	\hline
\end{tabular}\\ \\

Note that in \cite{suel2006, suel2007, cagatay2009}, the higher values $\Gamma_k = 25000$ and $\Gamma_s = 20$ (native), and $k_k = 5000$ and $k_m = 2500$ (SynEx) are used to establish molecule number. We use the lower values here to elucidate the effects of intrinsic noise. This assumption is relaxed in the next section, where we return to the high-number regime. In the main text, we discuss how the effects of noise are robust to this choice.

\subsection{Relaxing the model assumptions}
\slabel{robust}

To relax the first two simplifying assumptions made in the main text, we consider a stochastic model that has high molecular copy numbers, and that accounts explicitly for the mRNA dynamics and the dynamics of competitive degradation. The model follows from our earlier work \cite{cagatay2009}, and consists of a set of coupled chemical reactions for each circuit. Due to the complexity of the model, we do not solve the master equation explicitly, but rather we simulate the dynamics using the Gillespie algorithm \cite{gillespie1977}.

For the native circuit, the reactions and rates are:\\

\begin{tabular}{|l|l|}
	\hline
	Reaction & Rate \\
	\hline \hline
	P$^{\rm const}_{\rm ComK}$ $\to$ P$^{\rm const}_{\rm ComK}$ + mRNA$_{\rm ComK}$
		& $k_1$ (varied) \\
	P$_{\rm ComK}$ $\to$ P$_{\rm ComK}$ + mRNA$_{\rm ComK}$
		& $k_2 n^h/(k_k^h+n^h)$ \\
	mRNA$_{\rm ComK}$ $\to$ mRNA$_{\rm ComK}$ + ComK
		& $k_3 = 0.2$/s \\
	P$^{\rm const}_{\rm ComS}$ $\to$ P$^{\rm const}_{\rm ComS}$ + mRNA$_{\rm ComS}$
		& $k_4 = 0$/s \\
	P$_{\rm ComS}$ $\to$ P$_{\rm ComS}$ + mRNA$_{\rm ComS}$
		& $k_5/[1+(n/k_s)^p]$ \\
	mRNA$_{\rm ComS}$ $\to$ mRNA$_{\rm ComS}$ + ComS
		& $k_6 = 0.2$/s \\
	mRNA$_{\rm ComK}$ $\to$ $\emptyset$
		& $k_7 = 0.005$/s \\
	ComK $\to$ $\emptyset$
		& $k_8 = 10^{-4}$/s \\
	mRNA$_{\rm ComS}$ $\to$ $\emptyset$
		& $k_9 = 0.005$/s \\
	ComS $\to$ $\emptyset$
		& $k_{10} = 10^{-4}$/s \\
	MecA + ComK $\to$ MecA$_{\rm K}$
		& $k_{11}/\Omega$ \\
	MecA$_{\rm K}$ $\to$ MecA + ComK
		& $k_{-11} = 5\times 10^{-4}$/s \\
	MecA$_{\rm K}$ $\to$ MecA
		& $k_{12} = 0.05$/s \\
	MecA + ComS $\to$ MecA$_{\rm S}$
		& $k_{13}/\Omega$ \\
	MecA$_{\rm S}$ $\to$ MecA + ComS
		& $k_{-13} = 5\times 10^{-5}$/s \\
	MecA$_{\rm S}$ $\to$ MecA
		& $k_{14} = 4\times 10^{-5}$/s \\
	\hline
\end{tabular}
\begin{tabular}{|l|}
	\hline
	Additional parameters \\
	\hline \hline
	$k_2 = 0.19$/s \\
	$k_5 = 0.0015$/s \\
	$k_{11} = 2\times 10^{-6}$/s \\
	$k_{13} = 4.5\times 10^{-6}$/s \\
	$k_k = 5000$ \\
	$k_s = 833$ \\
	$h = 2$ \\
	$p = 5$ \\
	$\Omega = 1.66$ $\mu$m$^3$ \\
	$M_{\rm T} = 500$ \\
	\\
	\\
	\\
	\\
	\\
	\\
	\hline
\end{tabular}\\ \\

Here P$_{\rm gene}$ denotes the promoter of the corresponding gene (constitutive or regulated), and MecA$_{\rm K}$ and MecA$_{\rm S}$ represent the complex of MecA bound to ComK and ComS, respectively. As in the main text, $n$ is the number of ComK molecules per cell. $\Omega$ represents the cell volume, and $M_{\rm T}$ gives the total number of MecA molecules. Upon adiabatically eliminating the faster mRNA dynamics and the MecA dynamics, one obtains the model in the previous section, except with the larger molecule numbers $\Gamma_k = 25000$ and $\Gamma_s=20$ \cite{suel2007}. The control parameter here, $k_1$, is related to the control parameter in the reduced model, $\alpha_k$, via $k_1 = k_7\alpha_k/k_3$ \cite{suel2007}.

For the SynEx circuit, the reactions and rates are:\\

\begin{tabular}{|l|l|}
	\hline
	Reaction & Rate \\
	\hline \hline
	P$^{\rm const}_{\rm ComK}$ $\to$ P$^{\rm const}_{\rm ComK}$ + mRNA$_{\rm ComK}$
		& $k_1$ (varied) \\
	P$_{\rm ComK}$ $\to$ P$_{\rm ComK}$ + mRNA$_{\rm ComK}$
		& $k_2 n^h/(k_k^h+n^h)$ \\
	P$^{\rm const}_{\rm MecA}$ $\to$ P$^{\rm const}_{\rm MecA}$ + mRNA$_{\rm MecA}$
		& $k_3 = 0.0019$/s \\
	P$_{\rm MecA}$ $\to$ P$_{\rm MecA}$ + mRNA$_{\rm MecA}$
		& $k_4 n^p/(k_m^p+n^p)$ \\
	mRNA$_{\rm ComK}$ $\to$ mRNA$_{\rm ComK}$ + ComK
		& $k_5 = 0.2$/s \\
	mRNA$_{\rm MecA}$ $\to$ mRNA$_{\rm MecA}$ + MecA
		& $k_6 = 0.2$/s \\
	MecA + ComK $\to$ MecA
		& $k_7/\Omega$ \\
	mRNA$_{\rm ComK}$ $\to$ $\emptyset$
		& $k_8 = 0.005$/s \\
	mRNA$_{\rm MecA}$ $\to$ $\emptyset$
		& $k_9 = 0.005$/s \\
	MecA $\to$ $\emptyset$
		& $k_{10} = 10^{-4}$/s \\
	ComK $\to$ $\emptyset$
		& $k_{11} = 10^{-4}$/s \\
	\hline
\end{tabular}
\begin{tabular}{|l|}
	\hline
	Additional parameters \\
	\hline \hline
	$k_2 = 0.19$/s \\
	$k_4 = 0.0625$/s \\
	$k_7 = 4\times 10^{-8}$/s \\
	$k_k = 5000$ \\
	$k_m = 2500$ \\
	$h = 2$ \\
	$p = 2$ \\
	$\Omega = 1.66$ $\mu$m$^3$ \\
	\\
	\\
	\\
	\hline
\end{tabular}\\ \\

Once again, P$_{\rm gene}$ denotes the promoter of the corresponding gene, $n$ is the number of ComK molecules per cell, and $\Omega$ represents the cell volume. Upon adiabatically eliminating the faster mRNA dynamics, one obtains the model in the previous section, except with the larger molecule numbers $k_k = 5000$ and $k_m=2500$ \cite{cagatay2009}. The control parameter here, $k_1$, is related to the control parameter in the reduced model, $\alpha_k$, via $k_1 = k_8\alpha_k/k_5$.

To relax the third simplifying assumption made in the main text, we consider a stochastic model of the SynExSlow circuit. The model follows from our earlier work \cite{cagatay2009}, and is similar to the stochastic model in the main text, except that there are three species with dynamic molecule numbers: ComK ($n$), MecA ($m$), and ComS ($\ell$). The production rate functions are, respectively,
\begin{align}
\label{eq:r1}
g_n &= \alpha_k + \frac{\beta_k n^h}{k_k^h + n^h}, &
q_n &= \alpha_m + \frac{\beta_m n^p}{k_m^p + n^p}, &
y_n &= \alpha_s + \frac{\beta_s n^h}{k_s^h + n^h},
\end{align}
and the degradation rate functions are, respectively,
\begin{align}
\label{eq:r2}
r_{nm\ell} &= \frac{\delta_k m}{1 + n/\Gamma_k + \ell/\Gamma_s} + \lambda_k, &
s &= \lambda_m, &
u_{nm\ell} &= \frac{\delta_s m}{1 + n/\Gamma_k + \ell/\Gamma_s} + \lambda_s.
\end{align}
The parameter values are \cite{cagatay2009}:\\

\begin{tabular}{|ll|}
	\hline
	Parameters & \\
	\hline \hline
	$\alpha_k$ (varied) & $\beta_s = 0.5$/s \\
	$\alpha_m = 0.075$/s & $\Gamma_k = 25000$ \\
	$\alpha_s = 0.5$/s & $\Gamma_s = 20$ \\
	$\delta_k = \delta_s = 2\times 10^{-6}$/s & $k_k = 5000$ \\
	$\lambda_k = \lambda_m = \lambda_s = 10^{-4}$/s & $k_m = 2500$ \\
	$\beta_k = 7.5$/s & $k_s = 500$ \\
	$\beta_m = 2.5$/s & $h = p = 2$ \\
	\hline
\end{tabular}\\ \\

Note that these values correspond to the high-molecule-number regime of the native and SynEx models of the main text, and thus this model of the SynExSlow circuit also relaxes the first simplifying assumption of low molecule number.

Finally, we note that at these parameter values, the deterministic analog of this model predicts that as a function of the control parameter $\alpha_k$, the excitable regime transitions directly into a damped oscillatory (mono-stable) regime, where two of the three eigenvalues of the Jacobian matrix are complex, and all have negative real part. That is, there is no standard oscillatory regime. Therefore, we define a heuristic boundary $\alpha_k^{(2)} = 0.15$/s after which the damping is clearly evident within the first $24$ hours of the deterministic dynamics (see Fig.\ S7). The fact that the stochastic dynamics exhibit sustained oscillations in the absence of a deterministically oscillatory regime, even beyond this heuristic boundary (Fig.\ S7), indicates that the ability of noise to induce oscillations in the SynExSlow circuit is especially strong.

\subsection{Spectral solution to the master equation}
\slabel{spectral}

We write the master equation (\eref{me}) as
\beq
\label{eq:me1}
\frac{dp_{nm}}{dt} = -\left( \Lop_n[g_n,r_{nm}] + \Lop_m[q_n,s_{nm}] \right) p_{nm}.
\eeq
Here $-\Lop$ is the linear birth-death operator, whose action on the probability distribution $p$ is described by
\beq
\label{eq:bd}
-\Lop_n[g_n,r_n] p_n = g_{n-1} p_{n-1} + r_{n+1} (n+1) p_{n+1} - (g_n + r_n n) p_n,
\eeq
where in general on an operator (here, $\Lop$) we use a subscript to denote the sector ($n$ or $m$) on which it acts.  Explicitly, then, the master equation reads
\beqn
\frac{dp_{nm}}{dt} &=& g_{n-1} p_{n-1,m} + r_{n+1,m} (n+1) p_{n+1,m} - (g_n + r_{nm}n) p_{nm} \nonumber\\
&&	+  q_n p_{n,m-1} + s_{n,m+1} (m+1) p_{n,m+1} - (q_n + s_{nm}m) p_{nm}.
\eeqn

We will now derive the spectral decomposition of the master equation by introducing the generating function.  For intuition, we will first introduce the generating function in the context of the one-dimensional system described by Eqn.\ \ref{eq:bd}, then extend our results to the full master equation.

The generating function is an expansion in a complete set of states, indexed by molecule number, for which the probabilities provide the expansion coefficients.  Denoting the states abstractly as $\ket{n}$, the generating function is defined
\beq
\ket{G} \equiv \sum_n p_n \ket{n},
\eeq
where the sum runs from $0$ to $\infty$ (as do all sums hereafter unless otherwise specified).
Summing Eqn.\ \ref{eq:bd} against $\ket{n}$ yields
\beqn
-\Lop\ket{G} &=& \sum_n g_{n-1} p_{n-1} \ket{n} + \sum_n r_{n+1} (n+1) p_{n+1} \ket{n}
	- \sum_n g_n p_n \ket{n} - \sum_n r_n n p_n \ket{n} \\
\label{eq:bd2}
&=& \sum_n g_n p_n \ket{n+1} + \sum_n r_n n p_n \ket{n-1}
	- \sum_n g_n p_n \ket{n} - \sum_n r_n n p_n \ket{n},
\eeqn
where the second step shifts the sum without penalty (i) in the first term, since $p_{-1}=0$, and (ii) in the second term, since it vanishes for $n=0$.  Eqn.\ \ref{eq:bd2} benefits from the introduction of two additional sets of operators: (i) operators corresponding to the evaluation of the production and degradation functions at particular values of $n$,
\beqn
\g \ket{n} &\equiv& g_n \ket{n}, \\
\rhat \ket{n} &\equiv& r_n \ket{n},
\eeqn
and (ii) raising and lowering operators that correspond to the shifts in $n$ caused by birth and death,
\begin{align}
\label{eq:a1}
\ap \ket{n} &\equiv \ket{n+1}, &
\bra{n} \ap &= \bra{n-1}, \\
\label{eq:a2}
\am \ket{n} &\equiv n \ket{n-1}, &
\bra{n} \am &= n \bra{n+1}.
\end{align}
In Eqns.\ \ref{eq:a1}-\ref{eq:a2}, for completeness, we have also presented the operators' actions to the left, which derive from the orthonormality condition $\ip{n}{n'} = \delta_{nn'}$.
In terms of the above operators, Eqn.\ \ref{eq:bd2} becomes
\beqn
-\Lop\ket{G} &=& \sum_n p_n \ap \g \ket{n} + \sum_n p_n \am \rhat \ket{n}
	- \sum_n p_n \g \ket{n} - \sum_n p_n \ap \am \rhat \ket{n}, \\
&=& \left( \ap \g + \am \rhat - \g - \ap \am \rhat \right) \sum_n p_n \ket{n}, \\
\label{eq:bd3}
&=& - \left( \ap - 1 \right) \left( \am \rhat - \g \right) \ket{G}.
\eeqn
Eqn.\ \ref{eq:bd3} reveals the form of the birth-death operator in generating function space.

We now generalize Eqn.\ \ref{eq:bd3} to the two-dimensional master equation of the model.  Defining the two-dimensional generating function,
\beq
\ket{G} \equiv \sum_{nm} p_{nm} \ket{n,m},
\eeq
the master equation becomes
\beq
\label{eq:me2}
\ket{\dot{G}} = - \left[ \left( \ap_n - 1 \right) \left( \am_n \rhat_{nm} - \g_n \right)
	+ \left( \ap_m - 1 \right) \left( \am_m \s_{nm} - \q_n \right) \right] \ket{G},
\eeq
where dot denotes the time derivative, and as before the subscripts on operators denote the sectors on which they act.

The key insight of the spectral method is that a master equation such as Eqn.\ \ref{eq:me2} can be simplified significantly by expansion in a wisely chosen set of eigenfunctions.  Since Eqn.\ \ref{eq:me2} describes two coupled birth-death processes, we choose to expand in the eigenfunctions of two {\it uncoupled} birth-death processes -- that is, processes with {\it constant} production rates $\gb$ and $\qb$ and degradation rates $\rb$ and $\bar{s}$, respectively.  Again for intuition we begin in one dimension, for which the operator describing a constant-rate birth-death process is given by Eqn.\ \ref{eq:bd3}:
\beq
-\Lop \ket{G} = - \left( \ap - 1 \right) \left( \am \rb - \gb \right) \ket{G} = -\rb \bp \bm \ket{G}.
\eeq
In the second step we have defined the shifted raising and lowering operators
\beqn
\bp &\equiv& \ap - 1, \\
\bm &\equiv& \am - \gb/\rb.
\eeqn
The operator $\bp\bm$ is a number operator whose eigenvalues are integers, which we call $j$; thus the eigenvalue relation for the constant-rate birth-death operator is
\beq
-\Lop \ket{j} = -\rb j \ket{j}.
\eeq
Accordingly, $\bp$ and $\bm$ raise and lower the eigenstates as $\ap$ and $\am$ do the $\ket{n}$ states (Eqn.\ \ref{eq:a1}-\ref{eq:a2}):
\begin{align}
\bp \ket{j} &= \ket{j+1}, &
\bra{j} \bp &= \bra{j-1}, \\
\bm \ket{j} &= j \ket{j-1}, &
\bra{j} \bm &= j \bra{j+1}.
\end{align}
As we will see, the spectral method exploits the expansion of the generating function in these eigenstates $\ket{j}$.

Returning to two dimensions, it is clear that we would like to write the coupled master equation in terms of the uncoupled rates, which we do by adding them to and subtracting them from Eqn.\ \ref{eq:me2}:
\beqn
\ket{\dot{G}} &=& - \left[ \left( \ap_n - 1 \right) \left( \am_n [ \rhat_{nm} + \rb - \rb ] - \g_n - \gb + \gb \right)
	+ \left( \ap_m - 1 \right) \left( \am_m [ \s_{nm} + \bar{s} - \bar{s} ] - \q_n - \qb + \qb \right) \right] \ket{G} \qquad\\
&=& - \left[ \left( \ap_n - 1 \right) \left( \am_n \rb - \gb - \am_n \gam_{nm} + \Gam_n \right)
	+ \left( \ap_m - 1 \right) \left( \am_m \bar{s} - \qb - \am_m \lam_{nm} + \Lam_n \right) \right] \ket{G} \\
\label{eq:me3}
&=& - \left[ \rb \bp_n \bm_n - \bp_n \left( \bm_n + \frac{\gb}{\rb} \right) \gam_{nm} + \bp_n \Gam_n
+ \bar{s} \bp_m \bm_m - \bp_m \left( \bm_m + \frac{\qb}{\bar{s}} \right) \lam_{nm} + \bp_m \Lam_n \right] \ket{G}.
\eeqn
In the second step we define operators which capture the deviations between the constants and the coupled rates,
\begin{align}
\Gam_n &\equiv \gb - \g_n, &
\Lam_n &\equiv \qb - \q_n, \\
\gam_{nm} &\equiv \rb - \rhat_{nm}, &
\lam_{nm} &\equiv \bar{s} - \s_{nm},
\end{align}
and in the third step we write the raising and lowering operators in terms of their shifted counterparts:
\begin{align}
\label{eq:b1}
\bp_n &\equiv \ap_n - 1, &
\bp_m &\equiv \ap_m - 1, \\
\label{eq:b2}
\bm_n &\equiv \am_n - \gb/\rb, &
\bm_m &\equiv \am_m - \qb/\bar{s}.
\end{align}
We now expand the generating function in the eigenstates of $\bp_n\bm_n$ and $\bp_m\bm_m$, with eigenvalues $j$ and $k$, respectively,
\beq
\ket{G} = \sum_{jk} G_{jk} \ket{j,k},
\eeq
which makes Eqn.\ \ref{eq:me3}
\beq
\sum_{jk} \dot{G}_{jk} \ket{j,k} = - \sum_{jk} G_{jk} \left[ \rb j - \bp_n \bm_n \gam_{nm}
	- \frac{\gb}{\rb} \bp_n \gam_{nm} + \bp_n \Gam_n
	+ \bar{s} k - \bp_m \bm_m \lam_{nm}	
	- \frac{\qb}{\bar{s}} \bp_m \lam_{nm} + \bp_m \Lam_n \right] \ket{j,k}.
\eeq
Projecting from the left with the state $\bra{j,k}$,
\beqn
\dot{G}_{jk} &=& -(\rb j + \bar{s} k) G_{jk} - \sum_{j'} G_{j'k} \bra{j} \bp_n \Gam_n \ket{j'}
	- \sum_{j'k'} G_{j'k'} \bra{j,k} \bp_m \Lam_n \ket{j',k'} \nonumber\\
&&	+ \sum_{j'k'} G_{j'k'} \bra{j,k} \bp_n \bm_n \gam_{nm} \ket{j',k'}
	+ \frac{\gb}{\rb} \sum_{j'k'} G_{j'k'} \bra{j,k} \bp_n \gam_{nm} \ket{j',k'} \nonumber\\
&&	+ \sum_{j'k'} G_{j'k'} \bra{j,k} \bp_m \bm_m \lam_{nm} \ket{j',k'}
	+ \frac{\qb}{\bar{s}} \sum_{j'k'} G_{j'k'} \bra{j,k} \bp_m \lam_{nm} \ket{j',k'},
\eeqn
and noting the actions of $\bp$ and $\bm$ to the left (Eqns.\ \ref{eq:b1}-\ref{eq:b2}) yields the dynamics of the expansion coefficients,
\beq
\label{eq:me4}
\dot{G}_{jk} =  -(\rb j + \bar{s} k) G_{jk}
	- \sum_{j'} \left[ \Gamma_{j-1,j'} G_{j'k}
		+ \Lambda_{jj'} G_{j',k-1} \right]
	+ \sum_{j'k'} \left[ j \gamma_{jj'}^{kk'}
		+ \frac{\gb}{\rb} \gamma_{j-1,j'}^{kk'}
		+ k \lambda_{jj'}^{kk'}
		+ \frac{\qb}{\bar{s}} \lambda_{jj'}^{k-1.k'} \right] G_{j'k'},
\eeq
where the deviations, rotated into eigenspace, have become matrices and tensors:
\begin{align}
\label{eq:d1}
\Gamma_{jj'} &\equiv \bra{j} \Gam_n \ket{j'} = \sum_n \ip{j}{n} (\gb - g_n) \ip{n}{j'}, &
\Lambda_{jj'} &\equiv \bra{j} \Lam_n \ket{j'} = \sum_n \ip{j}{n} (\qb - q_n) \ip{n}{j'}, \\
\gamma_{jj'}^{kk'} &\equiv \bra{j,k} \gam_{nm} \ket{j',k'} &
\lambda_{jj'}^{kk'} &\equiv \bra{j,k} \lam_{nm} \ket{j',k'} \nonumber\\
\label{eq:d2}
&= \sum_{nm} \ip{j}{n} \ip{k}{m} (\rb - r_{nm}) \ip{n}{j'} \ip{m}{k'}, &
&= \sum_{nm} \ip{j}{n} \ip{k}{m} (\bar{s} - s_{nm}) \ip{n}{j'} \ip{m}{k'}.
\end{align}
With Eqn.\ \ref{eq:me4}, we have turned the original master equation (Eqn.\ \ref{eq:me2}) into a linear algebraic equation, involving only matrix and tensor multiplication.  In steady state Eqn.\ \ref{eq:me4} can be solved perturbatively by treating the first term on the right-hand side as larger than the others and iterating to convergence.  

Algorithmically, then, the procedure prescribed by the spectral method is the following.  First we compute the deviation matrices and tensors from the given rate functions $g_n$, $r_{nm}$, $q_n$, and $s_{nm}$, and our chosen `gauges' $\gb$, $\rb$, $\qb$, and $\bar{s}$.  Then we solve for the steady state of Eqn.\ \ref{eq:me4} by iteration to obtain the expansion coefficients $G_{jk}$.  Finally, we compute the probability distribution by taking the inverse transform:
\beq
p_{nm} = \sum_{jk} G_{jk} \ip{n}{j} \ip{m}{k}.
\eeq
The probability distribution provides the complete stochastic description of the steady-state process.
Note that both the deviation matrices/tensors and the probability distribution are computed via multiplication against the eigenmodes $\ip{n}{j}$ and $\ip{m}{k}$ or their conjugates $\ip{j}{n}$ and $\ip{k}{m}$; these are efficiently precomputed via recursive update rules.

For the SynEx circuit, the dynamics of the spectral expansion coefficients are simplified because we have $r_{nm} \rightarrow r_m$ and $s_{nm} \rightarrow \bar{s}$:
\beq
\label{eq:me7}
\dot{G}_{jk} =  -(\rb j + \bar{s} k) G_{jk}
	- \sum_{j'} \left[ \Gamma_{j-1,j'} G_{j'k}
		+ \Lambda_{jj'} G_{j',k-1} \right]
	+ \sum_{k'} \left[ j \gamma_{kk'} G_{jk'}
		+ \frac{\gb}{\rb} \gamma_{kk'} G_{j-1,k'} \right],
\eeq
where
\begin{align}
\label{eq:d3}
\Gamma_{jj'} &\equiv \sum_n \ip{j}{n} (\gb - g_n) \ip{n}{j'}, &
\Lambda_{jj'} &\equiv \sum_n \ip{j}{n} (\qb - q_n) \ip{n}{j'}, \\
\label{eq:d4}
\gamma_{kk'} &\equiv \sum_m \ip{k}{m} (\rb - r_m) \ip{m}{k'}.
\end{align}
We see that the sparser coupling of the degradation terms (compared to the native circuit) has left us with a linear algebraic equation involving only matrices, not tensors.  The solution follows the steps outlined above for the native circuit.

The spectral method is more efficient than other solution techniques by many orders of magnitude \cite{walczak2009}. However, we find that here it becomes numerically unstable at sufficiently high numbers. Therefore, at high copy numbers, we solve for the steady state of the original master equation (\eref{me1}) by iteration. This is less efficient but more numerically stable. Specifically, we tile the columns of $p_{nm}$ into one vector $P_i$, and write the master equation as
\beq
\frac{dP_i}{dt} = \sum_{i'} M_{ii'} P_{i'} = \sum_{i'} \left(-D_i \delta_{ii'} + N_{ii'}\right) P_{i'},
\eeq
where $M_{ii'}$ is the reshaped operator in \eref{me1}, and $-D_i$ and $N_{ii'}$ are its diagonal and non-diagonal components, respectively. In steady state, we have
\beq
P_i = D_i^{-1}\sum_{i'} N_{ii'} P_{i'},
\eeq
which we solve iteratively until convergence.

\subsection{Chromosomal alterations and antibiotic resistance}

The native strain was a variant from a previous study \cite{suel2007} and had the following chromosomal alterations and antibiotic resistance:\\

\begin{tabular}{|l|l|l|}
	\hline
	Locus & Construct & Antibiotic Resistance \\
	\hline \hline
	AmyE & $ P_{hyperspank}$ comK  & Spectinomycin \\
	\hline
	SacA & $ P_{comG}$ cfp, $P_{comS}$ yfp & Chloramphenicol\\
	\hline
\end{tabular}\\ \\

The SynExSlow strain was modified from \cite{cagatay2009} as described in the main text and had the following chromosomal alterations and antibiotic resistance:\\

\begin{tabular}{|l|l|l|}
	\hline
	Locus & Construct & Antibiotic Resistance \\
	\hline \hline
	AmyE & $ P_{hyperspank} comK $ & Spectinomycin \\
	\hline
	SacA & $ P_{comG Kbox1} mecA^{xp} $ & Chloramphenicol \\
	\hline
	$ \Delta srfA,comS $ &  $P_{comG} cfp $ & Neomycin / Kanamycin \\
	\hline
	GltA & $ P_{rpsD} comS $, $ P_{comG} comS $ & Phleomycin \\
	\hline
\end{tabular}\\ \\


\subsection{Supplementary figures}

\renewcommand{\thefigure}{A\arabic{figure}}
\setcounter{figure}{0}

\begin{figure}[h]
\centering
\includegraphics[width=.9\textwidth]{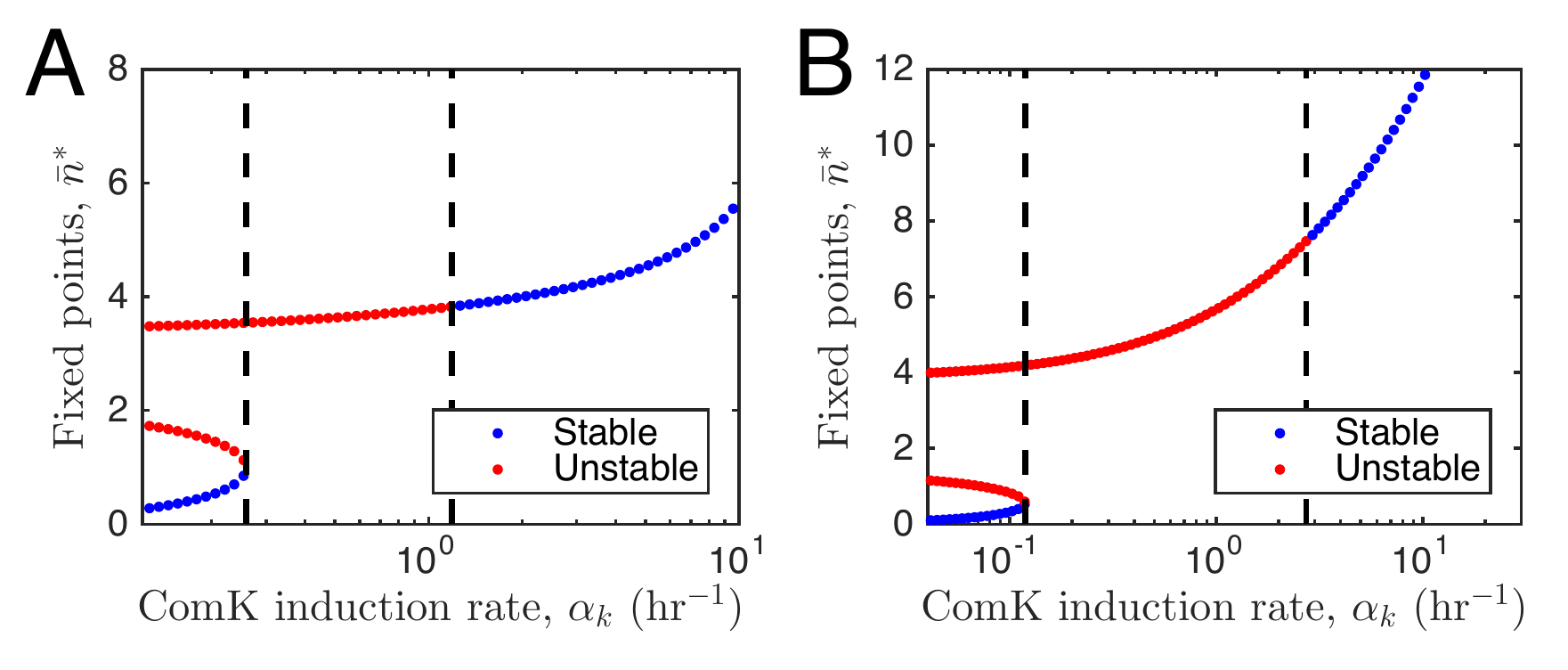}
\caption{{\bf Linear stability analysis of deterministic system predicts three dynamic regimes, for both the native and SynEx circuits.} Fixed points $\bar{n}^*$ satisfying the steady state of \erefs{ode1}{ode2} for (A) the native circuit and (B) the SynEx circuit. Each fixed point is stable if the real parts of the eigenvalues of the Jacobian matrix evaluated at that point are negative, and unstable otherwise. The Jacobian matrix is $J_{ij} = \partial F_i/\partial x_j$, where $x_1 \equiv \bar{n}$, $x_2 \equiv \bar{m}$, $F_1 \equiv d\bar{n}/dt$, and $F_2 \equiv d\bar{m}/dt$. For both circuits, there are three dynamic regimes. The excitable regime (low $\alpha_k$) is has three fixed points, one of which is stable. The oscillatory regime (intermediate $\alpha_k$) has one unstable fixed point. The mono-stable regime (high $\alpha_k$) has one stable fixed point. In the mono-stable regime, near the oscillatory regime, the eigenvalues are complex, indicating damped oscillations.}
\flabel{stability}
\end{figure}

\begin{figure}[h]
\centering
\includegraphics[width=\textwidth]{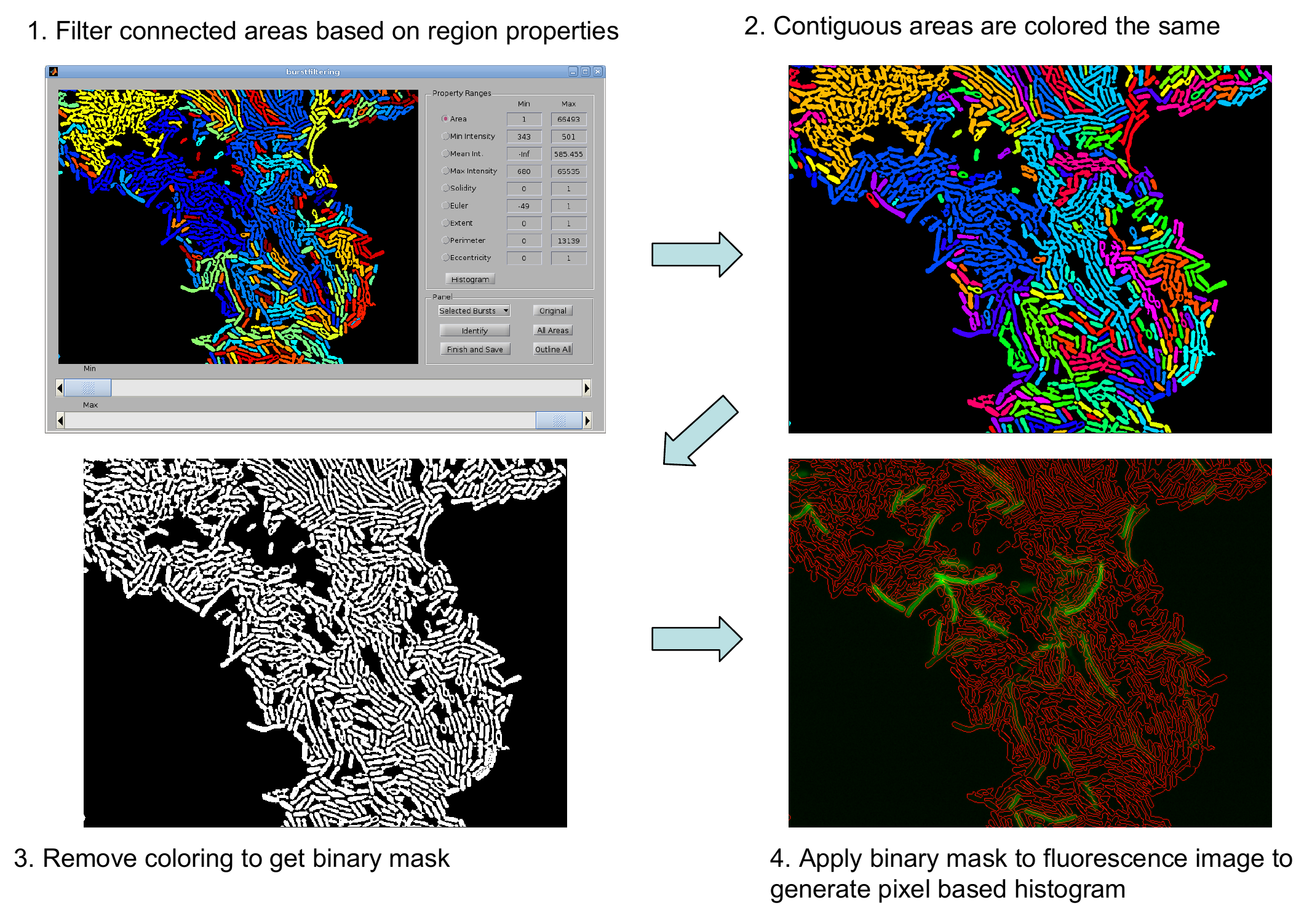}
\caption{{\bf Image analysis procedure.} Fluorescence histograms are generated via isolating cells from the background by identifying connected and contiguous areas, applying a binary mask, and binning the resulting pixel intensities. Only pixels that fall within cells, not the background, are included.}
\flabel{imaging}
\end{figure}

\begin{figure}[h]
\centering
\includegraphics[width=.6\textwidth]{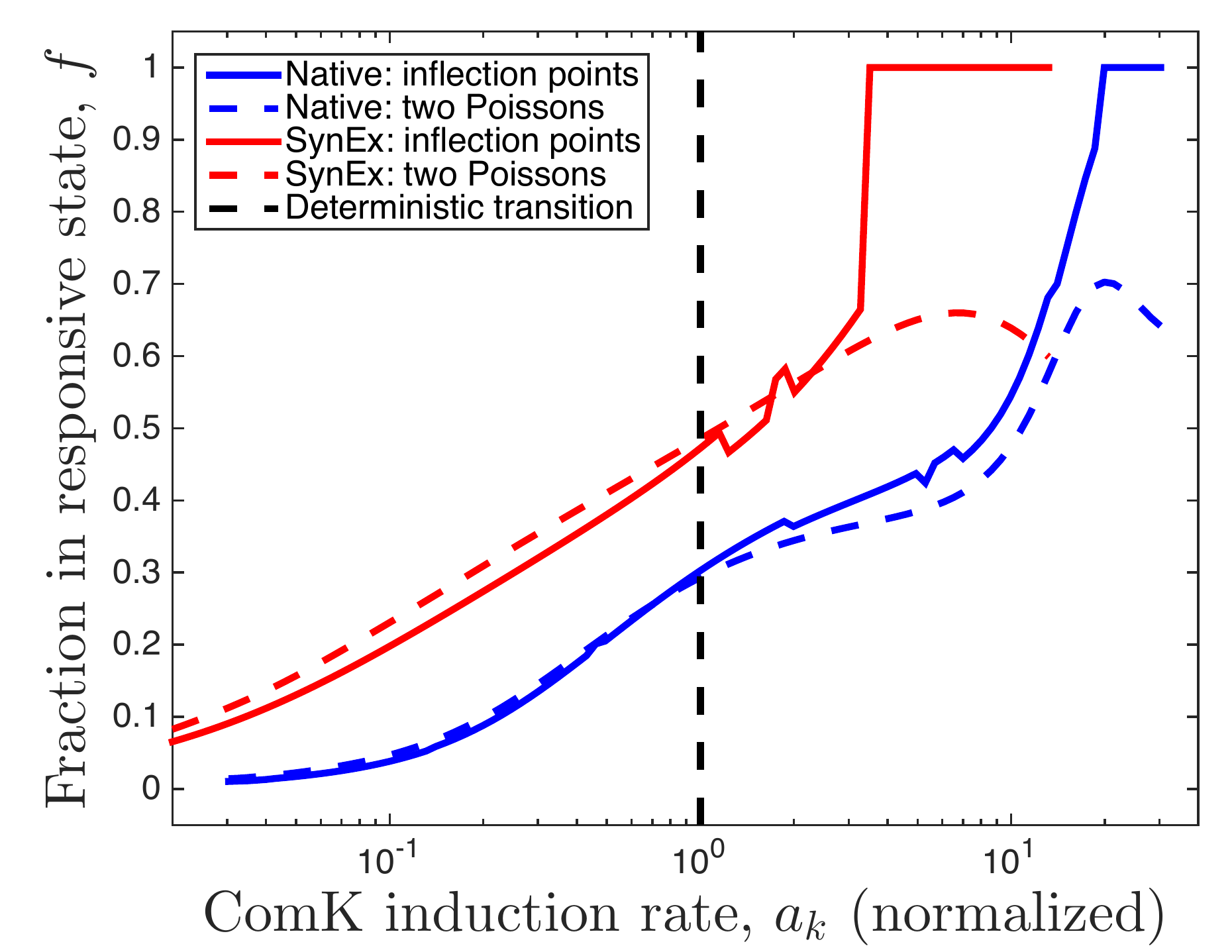}
\caption{{\bf Determination of high-response fraction $f$ is robust to calculation method.} Two independent methods are used to determine $f$: finding the inflection points, and fitting to a mixture of two Poisson distributions (see \sref{methods}). For both the native and SynEx circuit, we see that the two methods give results that correspond very closely to each other. The two-Poisson method gives smoother results, but does not capture the transition to $f = 1$, since a roughly equal mixture of two Poisson distributions with similar means ($f \sim 0.5$) will always provide a better fit than a single Poisson distribution ($f = 1$). Therefore, in \fref{theory}C and \fref{expt}C, we use the two-Poisson method for $f<1$ and use the inflection-point method to determine the transition to $f=1$.}
\flabel{fcalc}
\end{figure}

\begin{figure}[h]
\centering
\includegraphics[width=\textwidth]{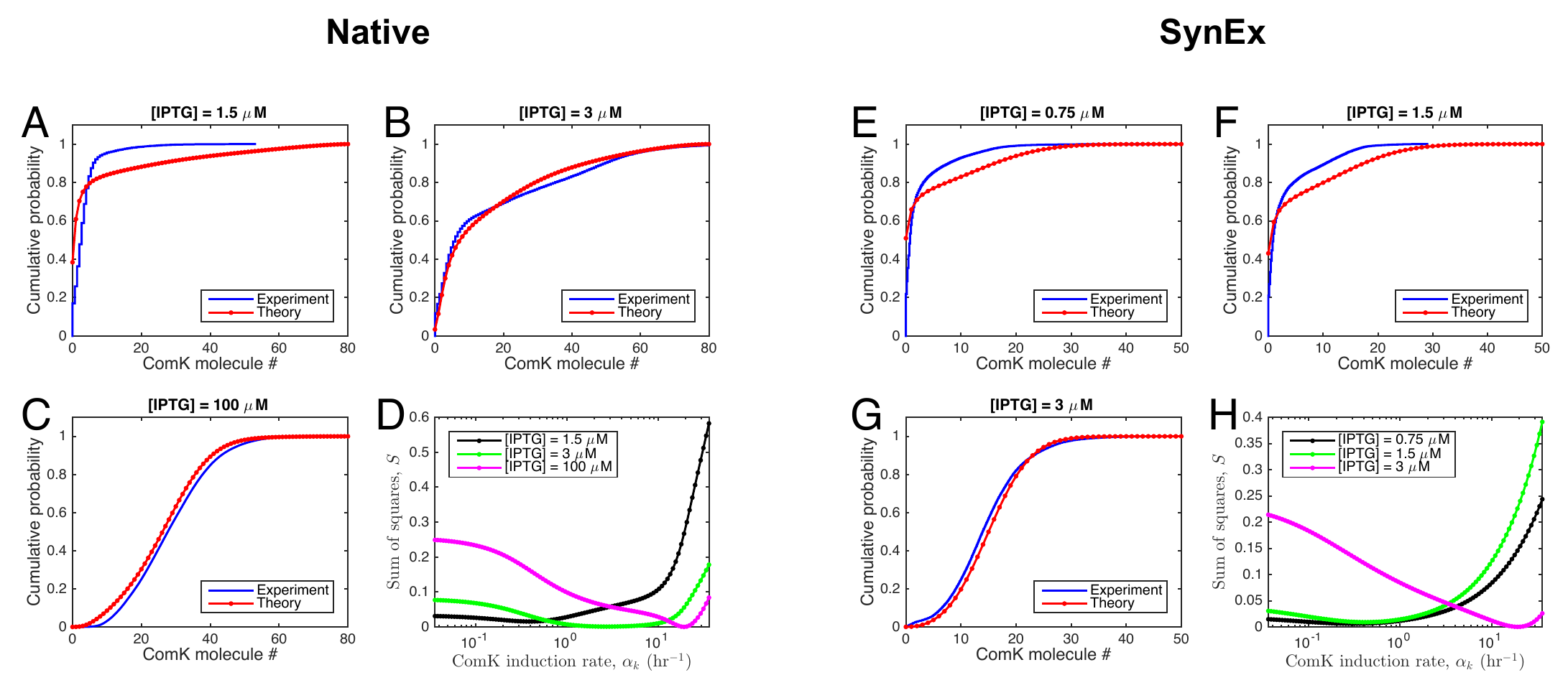}
\caption{{\bf Sample fits to experimental data and plots of sum-of-squares.} Cumulative probability distributions of fluorescence data are fit to the theoretical distributions by minimizing the sum of squared errors, for the (A-D) native and (E-H) SynEx circuits, as described in \sref{methods}.}
\flabel{fitting}
\end{figure}

\begin{figure}[h]
\centering
\includegraphics[width=\textwidth]{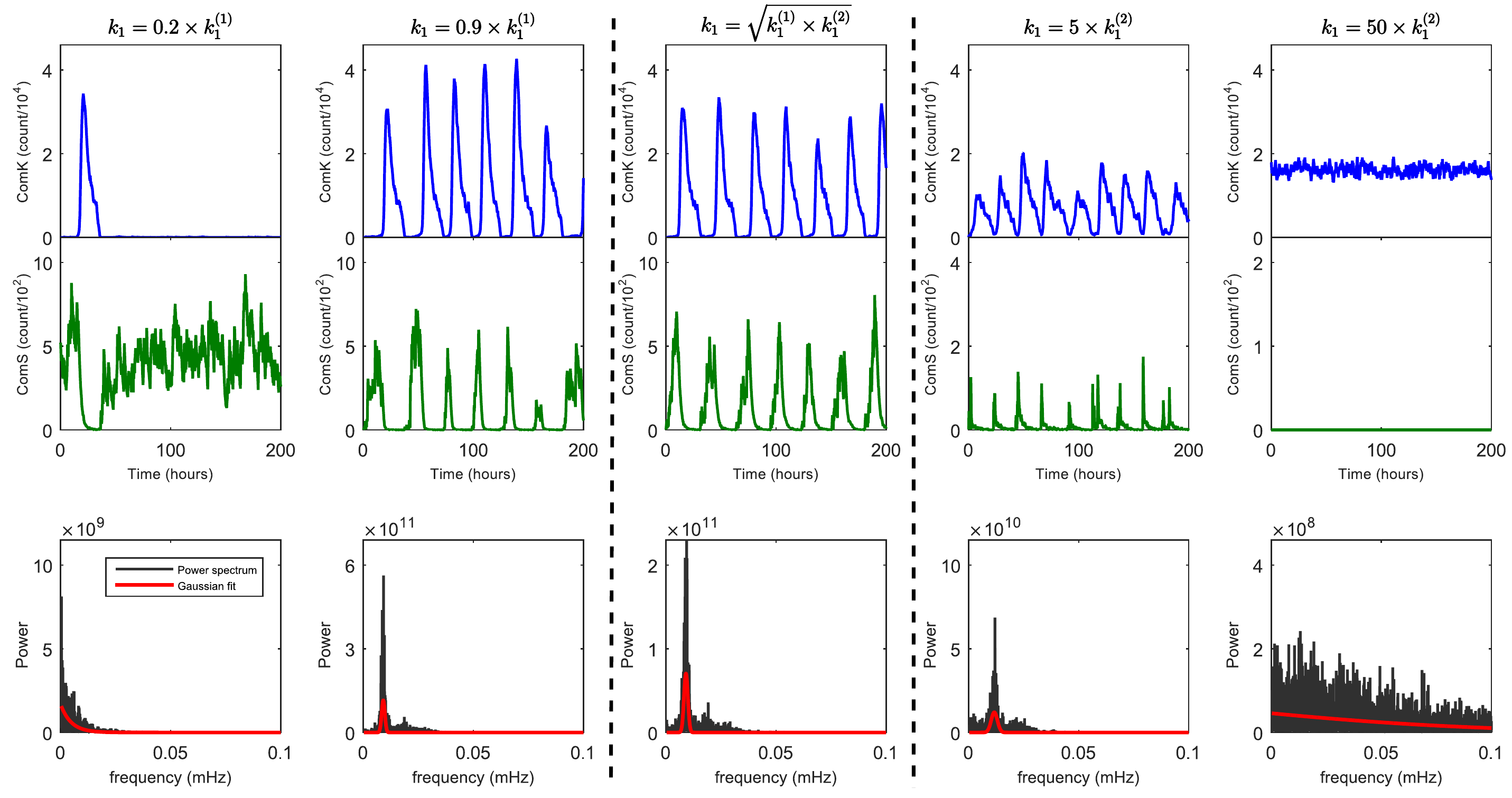}
\caption{{\bf Effects of noise persist when assumptions are relaxed in the model of the native circuit.} Top two rows show ComK and ComS time series from Gillespie simulations of the relaxed model that includes mRNA and competitive degradation dynamics (see \sref{robust}). Far from the deterministic boundaries of the control parameter, $k_1^{(1)}$ and $k_1^{(2)}$ (indicated by the dashed vertical lines), the dynamics are excitable, oscillatory, and mono-stable as predicted (columns $1$, $3$, and $5$, respectively). However, near the boundaries, but outside the oscillatory regime, noise causes oscillations to persist, due to either repeated excitations (column $2$) or prevention of damping (column $4$), confirming the effects seen in the reduced model of the main text. The persistence of oscillations is verified by computing the power spectrum $P(\omega) = |\tilde{n}(\omega)|^2$ from the Fourier transform of the ComK time series $n(t)$. For periodic signals, the power spectrum is peaked at a non-zero frequency $\omega$ (and in some cases its harmonics). Red line is a Gaussian fit to aid the eye.}
\flabel{native_full}
\end{figure}

\begin{figure}[h]
\centering
\includegraphics[width=\textwidth]{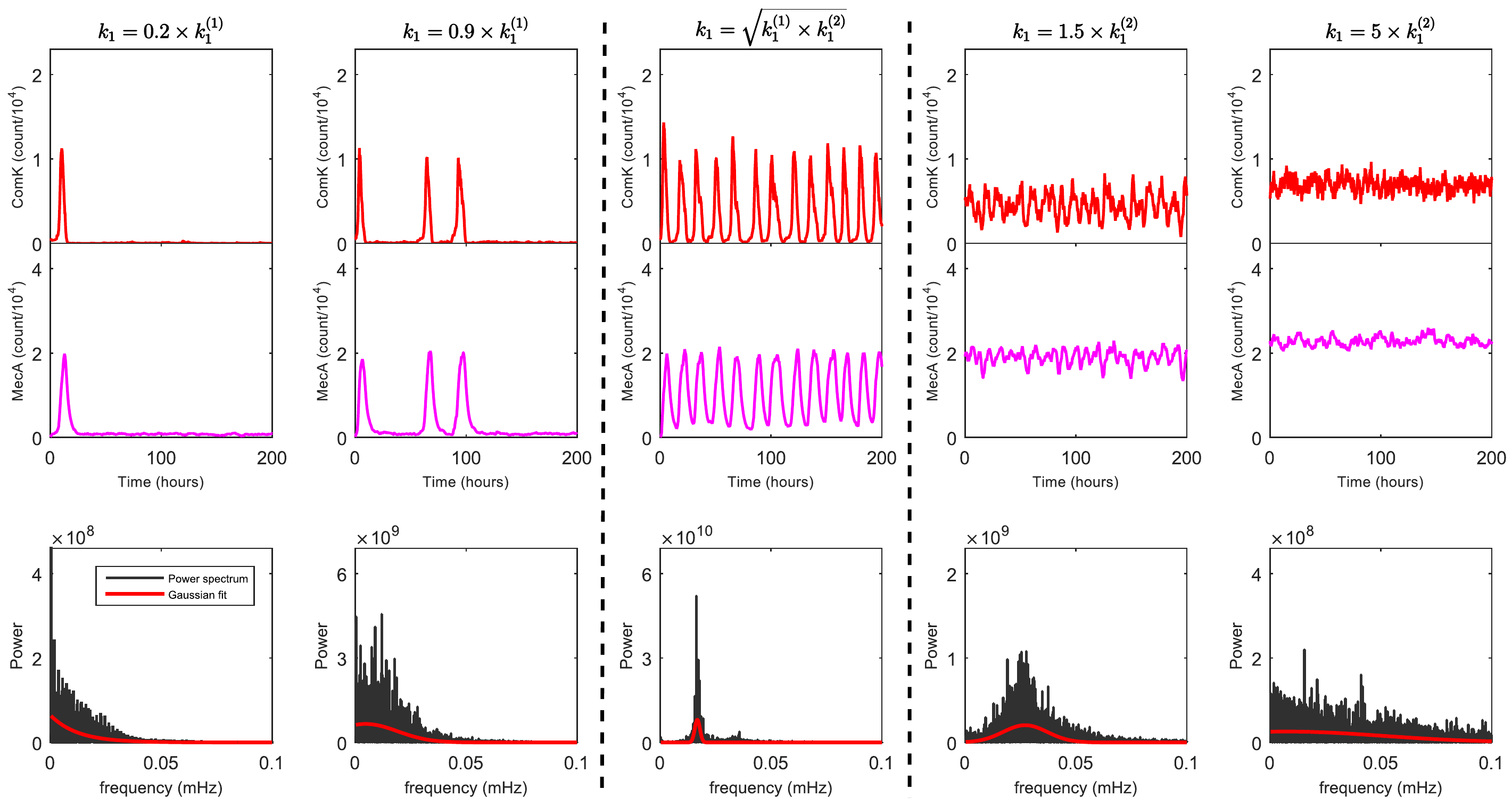}
\caption{{\bf Effects of noise persist when assumptions are relaxed in the model of the SynEx circuit.} As in \fref{native_full} but for the SynEx circuit. Once again, oscillations persist outside the deterministic boundaries as indicated by the peaked power spectra. Note, however, that oscillations are damped at the value $k_1 = 5k_1^{(2)}$ here, whereas in the native circuit they persist beyond this value (see \fref{native_full}). This confirms the effect seen in the main text that the prevention of damping is more pronounced in the native circuit than in the SynEx circuit.}
\flabel{synex_full}
\end{figure}

\begin{figure}[h]
\centering
\includegraphics[width=\textwidth]{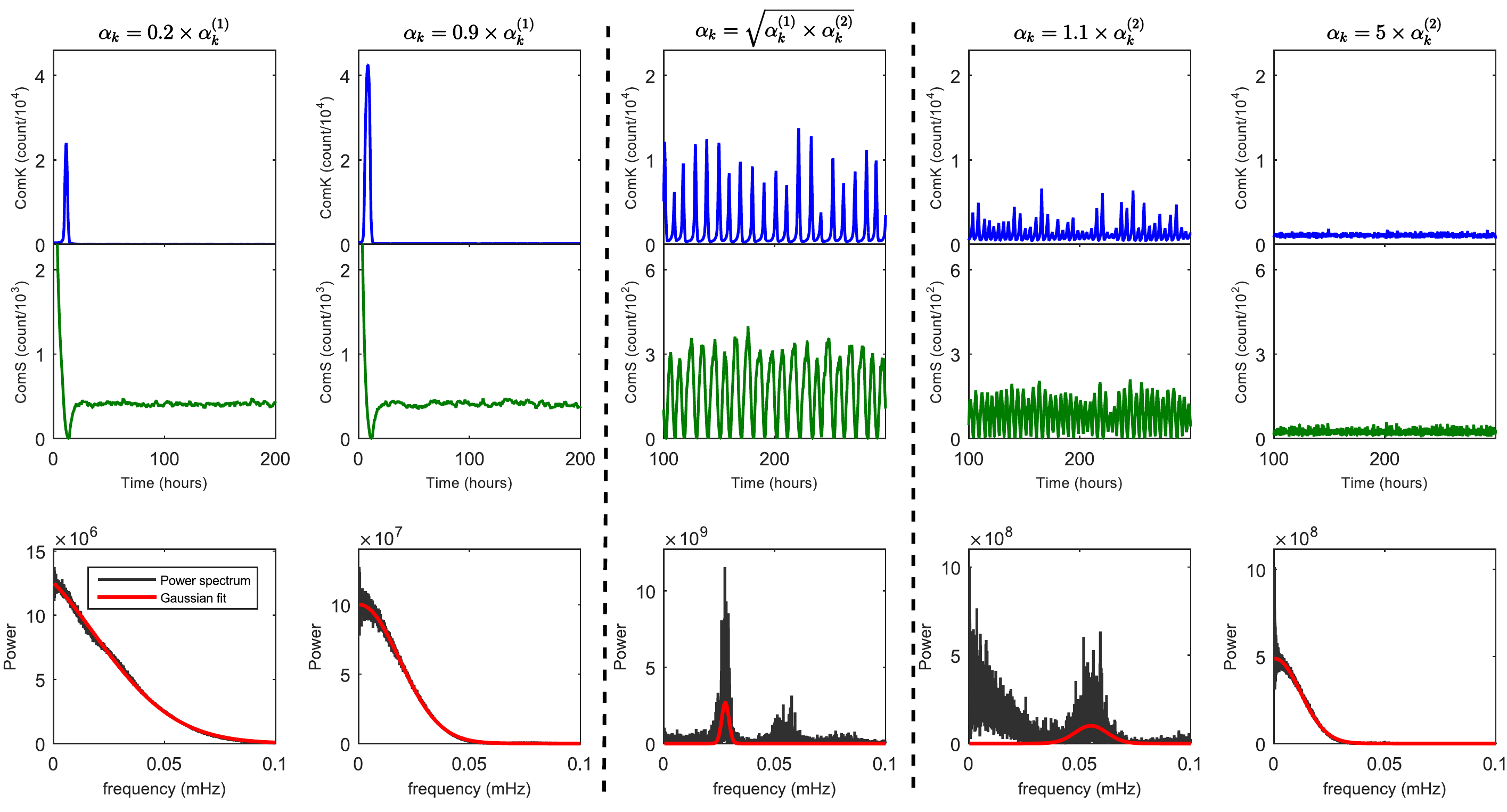}
\caption{{\bf Effects of noise diminish when molecule number is raised in the adiabatic model of the native circuit.} Gillespie simulations of the adiabatically reduced model of the native circuit (as in \fref{sim}), but for high molecule numbers ($\Gamma_k = 25000$ and $\Gamma_s = 20$). We see that $10\%$ outside the deterministically oscillatory regime, the stochastic dynamics are either non-oscillatory (column $2$) or weakly oscillatory (column $4$). This is in contrast to the low-molecule-number regime (\fref{sim}), where oscillations persist in these regions and beyond. We conclude that raising molecule number in the adiabatic model of the native circuit reduces the stochastic behavior to the deterministic behavior.}
\flabel{native_high}
\end{figure}

\begin{figure}[h]
\centering
\includegraphics[width=\textwidth]{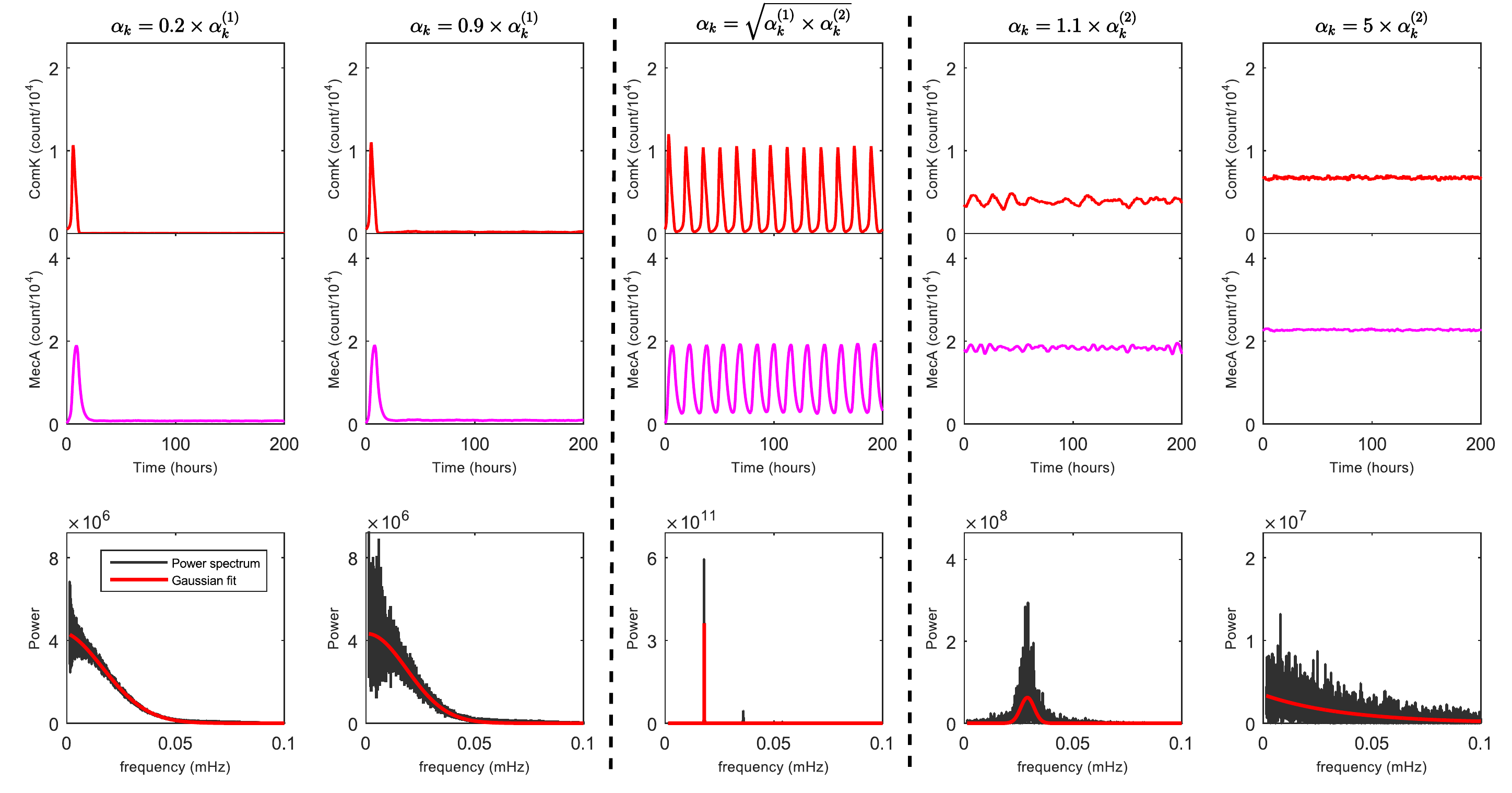}
\caption{{\bf Effects of noise diminish when molecule number is raised in the adiabatic model of the SynEx circuit.} As in \fref{native_high}, but for the adiabatically reduced model of the SynEx circuit at high molecule numbers ($k_k = 5000$ and $k_m = 2500$). Comparing to \fref{sim}, we similarly conclude that raising molecule number in the adiabatic model of the SynEx circuit reduces the stochastic behavior to the deterministic behavior.}
\flabel{synex_high}
\end{figure}

\begin{figure}[h]
\centering
\includegraphics[width=\textwidth]{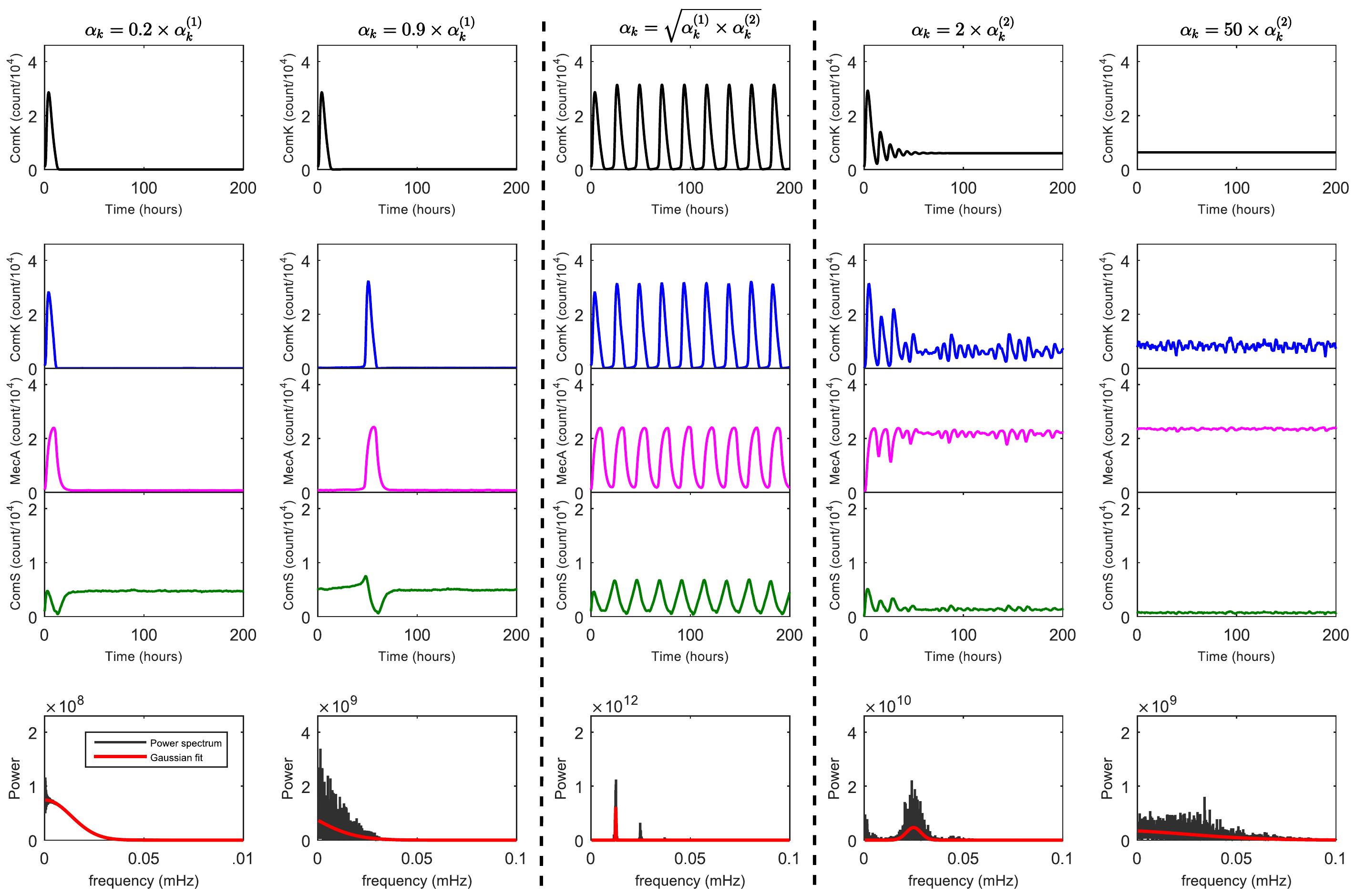}
\caption{{\bf Effects of noise persist in a model of the SynExSlow circuit.} Top row shows the deterministic ComK time series from the SynExSlow model (see \sref{robust}), while the next three rows show the stochastic ComK, MecA, and ComS time series for the same model. Although the SynExSlow model only exhibits a damped oscillatory regime at these parameters, not a standard oscillatory regime (see \sref{robust}), we define a heuristic boundary $\alpha_k^{(2)} = 0.15$/s below which oscillations are not appreciably damped within the first 24 hours (column $3$), and above which they are (column $4$). We see that, as in \fref{synex_full}, noise prevents damping at large values of the control parameter, even at high molecule numbers (column $4$). However, as in \fref{synex_high}, noise does not induce repeated excitations at small values of the control parameter (column $2$). We conclude that the former effect is more robust.}
\flabel{synexslow}
\end{figure}

\end{document}